\RequirePackage{fix-cm}
\documentclass[smallcondensed]{svjour3}     
\smartqed  
\usepackage{longtable}
\usepackage{color}

\usepackage{graphicx}
\usepackage[pdftex]{thumbpdf}
\usepackage[
    colorlinks,bookmarksopen,bookmarksnumbered,citecolor=red,urlcolor=red,
    bookmarks=true,bookmarksnumbered=true,hypertexnames=false,
    breaklinks=true,linkbordercolor={1 1 1},pdfborder={0 0 0}]{hyperref}
\hypersetup{
    pdfauthor={Marco Kuhrmann},
    pdftitle={On the Pragmatic Design of Literature Studies in Software Engineering: An Experience-based Guideline},
    pdfsubject={Empirical Software Engineering},
    pdfkeywords={Systematic Literature Review, Systematic Mapping Study, Empirical Software Engineering, Method Proposal, SLR}
}

\usepackage{amsmath}
\usepackage{amssymb}
\usepackage{multirow}
\usepackage{array}
\usepackage[export]{adjustbox}

\usepackage{pifont}
\newcommand{\cmark}{\ding{51}}%
\newcommand{\starmark}{\ding{72}}%
\newcommand{\emptystarmark}{\ding{73}}%

\def\BibTeX{{\rm B\kern-.05em{\sc i\kern-.025em b}\kern-.08em
    T\kern-.1667em\lower.7ex\hbox{E}\kern-.125emX}}
    
\spnewtheorem{artifact}{Artifact}{\bf}{\rm}
\spnewtheorem{synonym}{Synonym}{\bf}{\rm}

\begin{document}

\title{On the Pragmatic Design of Literature Studies in Software Engineering: An Experience-based Guideline
}

\titlerunning{On the Pragmatic Design of Literature Studies in Software Engineering}        

\author{Marco Kuhrmann \and Daniel M\'{e}ndez Fern\'{a}ndez \and Maya Daneva}


\institute{M. Kuhrmann \at
              University of Southern Denmark,\\
              M\ae rsk Mc-Kinney M\o ller Institute, Section Software Engineering,  \\
              Campusvej 55, 5230 Odense M, Denmark\\
              Tel.: +45 2460 1422\\
              \email{kuhrmann@acm.org}           
           \and
           D. M\'{e}ndez Fern\'{a}ndez \at
              Technical University of Munich, \\
              Institute for Informatics, Software \& Systems Engineering \\
              Boltzmannstr. 4, 85748 Garching, Germany \\
               Tel.: +49 89 289 17056\\
              \email{daniel.mendez@tum.de}
               \and
           M. Daneva \at
              University of Twente, \\
              Drinerlolaan 5, 7522 AE, Enschede The Netherlands\\
              Tel.: +31 53 4892889\\
              \email{m.daneva@utwente.nl}
}

\date{Received: date / Accepted: date}

\maketitle

\begin{abstract}
Systematic literature studies have received much attention in empirical software engineering in recent years. They have become a powerful tool to collect and structure reported knowledge in a systematic and reproducible way. We distinguish systematic literature reviews to systematically analyze reported evidence in depth, and systematic mapping studies to structure a field of interest in a broader, usually quantified manner. Due to the rapidly increasing body of knowledge in software engineering, researchers who want to capture the published work in a domain often face an extensive amount of publications, which need to be screened, rated for relevance, classified, and eventually analyzed. Although there are several guidelines to conduct literature studies, they do not yet help researchers coping with the specific difficulties encountered in the practical application of these guidelines. In this article, we present an experience-based guideline to aid researchers in designing systematic literature studies with special emphasis on the data collection and selection procedures. Our guideline aims at providing a blueprint for a practical and pragmatic path through the plethora of currently available practices and deliverables capturing the dependencies among the single steps. The guideline emerges from various mapping studies and literature reviews conducted by the authors and provides recommendations for the general study design, data collection, and study selection procedures. Finally, we share our experiences and lessons learned in applying the different practices of the proposed guideline.

\end{abstract}

\section{Introduction}
\label{sec:Introduction}
Systematic literature studies have received much attention in recent years as a powerful instrument to gather and structure reported knowledge in a systematic and reproducible way. We distinguish two types of secondary studies: 
\begin{description}
	\item[A \emph{Systematic Mapping Study}] (SMS; Petersen et al.\ \cite{PFMM08}) is a method to build a classification schema for topics studied in a field of interest. By counting the number of publications for categories within a schema, the coverage and maturity of the research field can be determined. Graphical maps showing the number of publications in the different categories of the schema represent the study results. Mapping studies usually cover a broader range of publications as the analysis focuses on the key terms and abstracts of publications.
		\item[A \emph{Systematic Literature Review}] (SLR; also: Systematic Review, SR; Kitchenham et al.\ \cite{Kitchenham:2015rt}) is a means to identify, analyze and interpret reported evidence related to a set of specific research questions in a way that is unbiased and (to a degree) repeatable. In contrast to mapping studies, systematic reviews usually cover a smaller, more specific range of publications while the analysis focuses on the details of the published contributions.
\end{description}
A mapping study is therefore often used to provide (and visualize) a big picture of a publication space while the systematic review is additionally concerned with analyzing and integrating the knowledge contained in the reviewed publications, as well as identifying inconsistencies among results, and areas that need more investigation.
Both types of secondary studies (also applicable in combination) allow to share a structured overview of the publications in a specific research area and a common understanding of the state of reported evidence in topics along a given (or emerging) classification scheme. Since the initially proposed guidelines to conduct literature studies in software engineering \cite{Kitchenham:2004fk}, we, as a community, could collect and systematize the procedures required, and we could see a boost of secondary studies in the various international evidence-based software engineering venues. This indicates the value of such studies to the research communities.

\paragraph{Problem Statement}
Since researchers face a variety of challenges for which available guidelines do not yet give sufficient practical advice; they either comprise generic workflows or provide methods and techniques in a compendium-like style \cite{Kitchenham:2015rt,PVK15}, or elaborate selected methods only, e.g., the effectiveness of certain selection procedures \cite{Ali:2014:ESS:2652524.2652557,Zhang:2011:IRS:1968229.1968314}. Hence, conducting a literature study still depends to a large extent on the expertise of the involved researchers. Furthermore, conducting literature studies, to a large extent, still lacks  tool support \cite{Hassler2016122,6681353,Tell:2016rz} thus making the research process as such difficult to implement; notably for novices. While working on a number of literature studies ourselves (Sect.~\ref{sec:ExamplesAndLessons}), we experienced the following challenges to be the most critical ones worth deeper examination:
\begin{itemize}
	\item How do we begin a secondary study, how do we build search strings adequate for given databases, and how can we control accurate results given the dependency to the expertise, experiences, and potential subconscious bias of the researchers?
	\item How do we deal with a large amount of data including hundreds or even thousands of potentially relevant papers to classify and structure, and how do we efficiently filter relevant results from irrelevant ones?
	\item How do we efficiently work in a distributed team? Which tools can we use to organize our (potentially distributed) way of working?
\end{itemize}
We experienced those challenges to concern mainly the design of a study \cite{Kitchenham:2015rt} and the data collection and study selection itself \cite{Zhang:2011:IRS:1968229.1968314}, notably independent of whether it is conducted as a systematic review or a mapping study. The choice of one particular study approach or a combination thereof (as for instance found in  \cite{PVK15} oftentimes) affects subsequent data analysis where the data is structured, classified, coded, and analyzed to draw conclusions in tune with the research questions. 

Despite the criticality of the initial design and data collection steps, little practical advice is given on how to effectively cope with the mentioned challenges. Existing guidelines are either too generic~\cite{Staples:2007:EUS:1282869.1282969}, or they focus on \emph{what} a design should accomplish rather than on \emph{how} and \emph{why} particular practices should be executed in a cost-effective way, and how these practices are interconnected with each other (see also our discussion in Sect.~\ref{sec:RelatedWork}). In turn, for each literature study, researchers need to carefully design and outline the process from the beginning again and again, and they need to work out or even re-invent their own set of best practices.

\paragraph{Contribution}
\label{sec:Contribution}
In this article, we report on our own experiences in conducting systematic literature studies and contribute
\begin{itemize}
	\item A detailed blueprint for the design, data collection, and study selection procedures steered by the aforementioned challenges.
	\item A set of practical lessons learned and supporting material readily available for use by other researchers approaching their own systematic literature studies.
\end{itemize}
We aim at supporting researchers, who already have a basic knowledge about the general guidelines, in their literature studies by providing a practical and pragmatic, experienced-based path through the available practices and deliverables capturing the dependencies among the single steps (Sect.~\ref{sec:RelatedWork}). Researchers can directly reuse our blueprint to design and conduct their own domain-specific literature study and build on top the data analysis to answer their individual research questions.

\paragraph{Outline}
\label{sec:Outline}
The remainder of this article is organized as follows: In Sect.~\ref{sec:Approach}, we present our experienced-based approach to design and set up a literature study. We describe our procedures as they emerged from our previously conducted studies. We also outline the handover to the data analysis, which depends on the type of the respective study (mapping study and/or systematic review) and the research questions previously defined. Our previously conducted studies from which we distill the blueprint are discussed in Sect.~\ref{sec:ExamplesAndLessons} along practical lessons we learned while conducting these studies. In Sect.~\ref{sec:RelatedWork}, we finally discuss related work and position our guideline, before concluding our article in Sect.~\ref{sec:Conclusion}. In the articles's appendix, we provide exemplary integrated workflows describing reusable standard workflows, and further complementing material.

\section{Study Design and Data Collection: An Experience-based Approach}
\label{sec:Approach}
\begin{figure*}[!ht]
  \includegraphics[width=\textwidth, right]{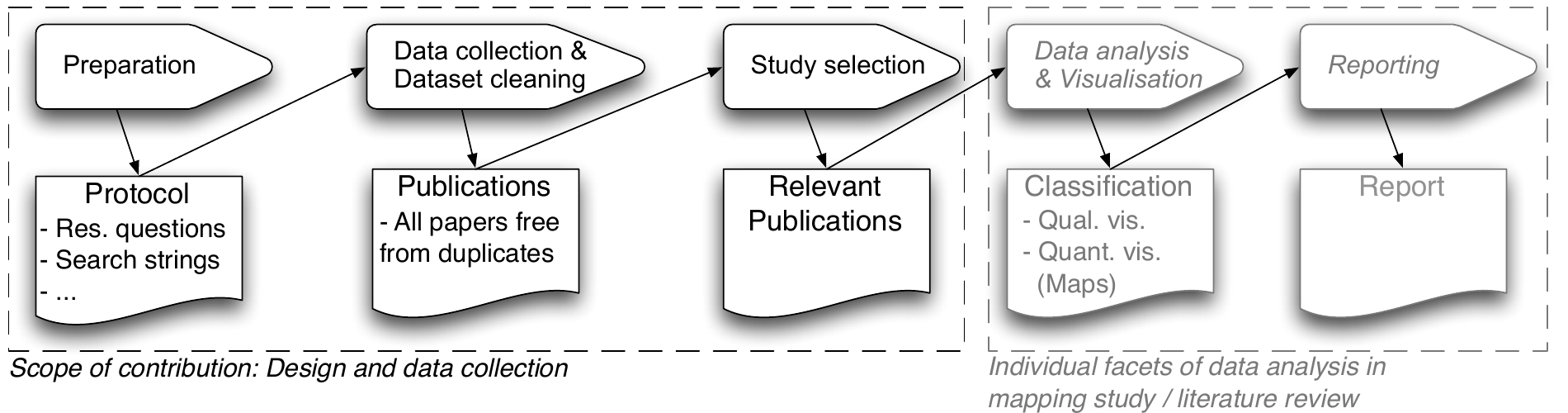}
  \caption{Overview of the presented approach and scoping.}
  \label{fig:01:ApproachOverview}
\end{figure*}

We provide an experienced-based guideline to support the study design, and to perform the data collection, cleaning, and study selection procedures. For each step, we provide a guideline complemented with small inline examples.
The guideline is organized in the three phases \emph{Preparation}, \emph{Data collection and Dataset cleaning}, and \emph{Study selection}. Figure~\ref{fig:01:ApproachOverview} provides a big picture of the whole process including the most important inputs and outputs for the respective phases. The figure also outlines the variations in the data analysis procedures that depend (in more detail) on whether it is a mapping study or a systematic review. 

Our guideline presented in this article emphasizes the early stages of a literature study and constitutes a new building block in the methodical instrumentation of evidence-based software engineering \cite{Kitchenham:2015rt}. A detailed discussion on the relation to existing guidelines and publications is provided in the related work in Sect.~\ref{sec:RelatedWork}. 

\subsection{Preparation}
\label{sec:Approach:Preparation}
The study preparation phase serves the purpose of setting up the study design including, inter alia, the definition of appropriate research questions, the choice of relevant literature databases, or the development of search queries. This phase relates to the \emph{planning} step mentioned in \cite{Kitchenham:2015rt} where, for instance, the protocol development is described. To set the scope of the search, inclusion and exclusion criteria need to be carefully outlined, and, if necessary, preliminary studies can be carried out to, among other things, support search string development or testing and improving the study design (see also test-retest procedures as mentioned in \cite{Kitchenham:2015rt}, or the quasi-gold standard search approach from \cite{Zhang:2011:IRS:1968229.1968314}). In the following, we describe the individual and minimum steps to be carried out during the preparation of a literature study and give examples. 

\subsubsection{Research Goals and Research Questions}
\label{sec:Approach:Preparation:GoalsAndRQs}
There is no silver bullet to define the goals of a literature study, as this strongly depends on the purpose of the study. In general, the primary goal of literature studies is to systematically collect reported knowledge in an area of interest. This can be done \emph{in-breadth}, usually in scope of mapping studies \cite{PFMM08} that quantify selected aspects reported in literature, or \emph{in-depth}, usually in scope of systematic reviews \cite{Kitchenham:2015rt} to analyze publications in detail. The purpose of a study eventually dictates the goals of the study, such as providing an overview of all relevant contributions dealing with a particular topic. 

Independent of the respective goals, we have found some general research questions particularly worth considering in a literature study, as they help elaborating a big picture and providing relevant background information about the publication space. Table~\ref{tab:StandardResearchQuestions} summarizes such generic research questions, which could be answered in every literature study---regardless of the particular study's scope and selected topic. 
\begin{table*}[!ht]
\centering
\caption{Exemplary standard research questions for literature studies.}
\label{tab:StandardResearchQuestions}
\begin{tabular}{@{}l@{\quad}p{0.92\textwidth}@{}}
    \hline\noalign{\smallskip}
    	\textbf{No.} & \textbf{Research Question}  \\
    	\noalign{\smallskip}\hline\hline\noalign{\smallskip}
	1    & Which/how many publications on [topic] are published? \\
	2    & Which/how many publications on [topic] are published over the years? \\
	3    & What is the scientific maturity of the publication set? \\
	4    & What is the contribution of the publication set? \\
	5    & What are observable mainstreams in the publication set? \\
	6    & What new approaches for [topic] are available? \\
\noalign{\smallskip}\hline\noalign{\smallskip}
\end{tabular}
\end{table*}

\begin{figure*}[t]
  \includegraphics[width=0.985\textwidth, right]{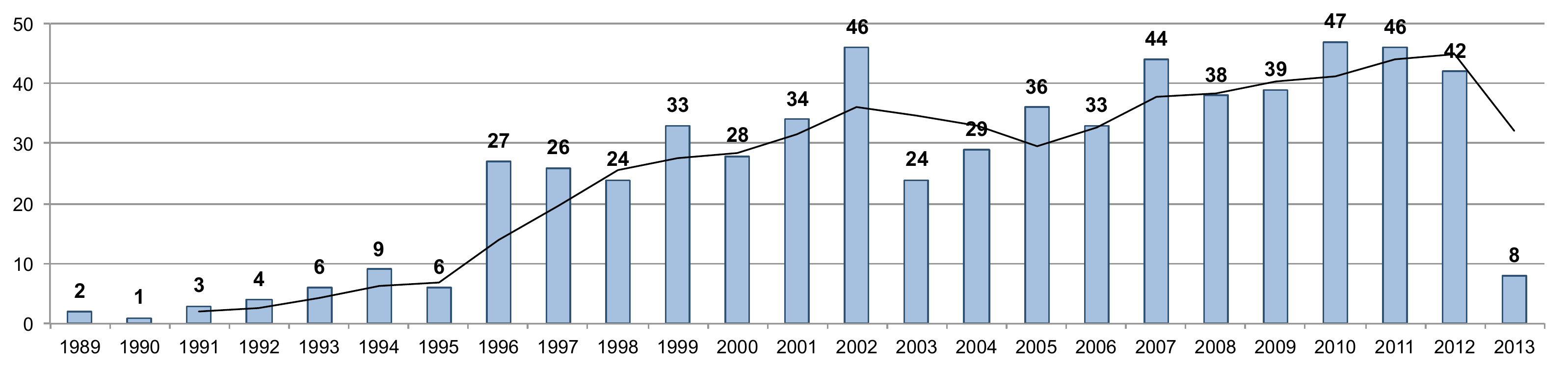}\\
  \includegraphics[width=\textwidth, right]{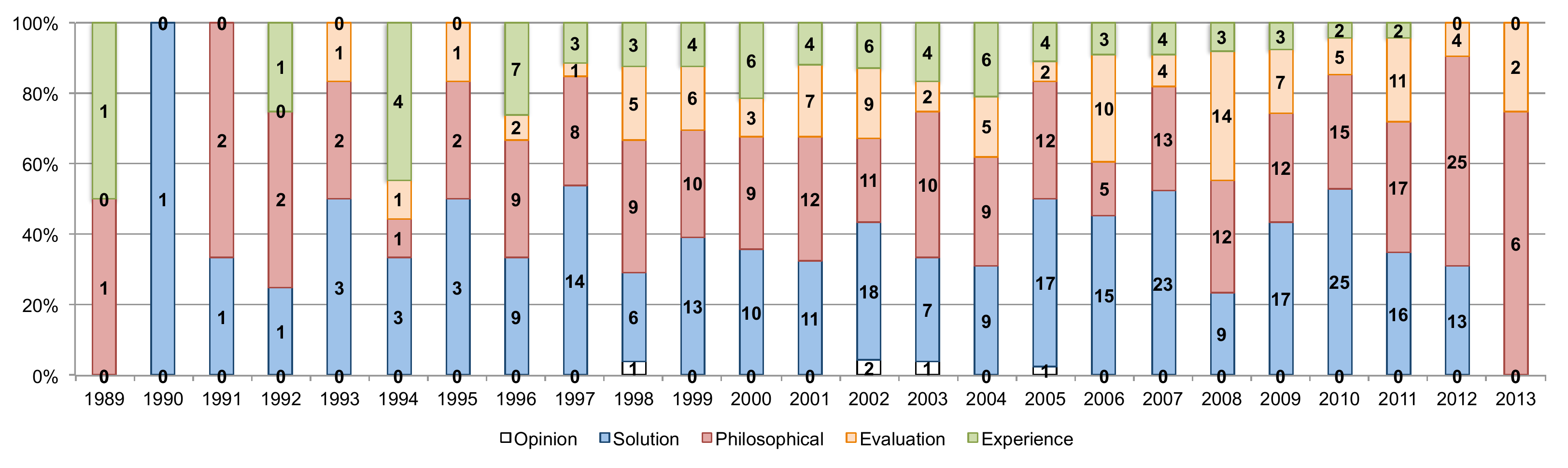}
  \caption{Exemplary demographic distribution of publications over specific facets (addressing research questions 1, 2, and 4) as taken from \cite{Kuhrmann:2015ix}. This figure illustrates on top the number of publications over time and per year depicting publication trends. The bottom part indicates to the maturity of the result set by providing information about the research type facets.}
  \label{fig:PublicationFrequencySample}
\end{figure*}

The research questions in Table~\ref{tab:StandardResearchQuestions} address the general descriptive aspects present in every result set. Questions 1 and 2 aim at drawing a demographic picture to outline the current state of a field under investigation, i.e., providing information about publication quantity and frequency. This information can be instrumented to show the development over time of the studied domain and to analyze trends, for example, an emerging or a maturing domain (as exemplarily depicted in Figure~\ref{fig:PublicationFrequencySample}). The level of detail and data type (quantitative or qualitative) further depends on the respective study type\footnote{Note that finding the ``right'' research question is a challenge and highly depends on the actual study type. For instance, Kitchenham et al.\ \cite{Kitchenham:2015rt} mention (standard) research questions for systematic reviews usually addressing the evaluation of impact and/effectiveness of certain paradigms, while mapping studies usually address more high-level questions with the purpose of providing some sort of categorization. The questions presented in Table~\ref{tab:StandardResearchQuestions} are addressing more the latter aspect, as this covers information available from all sorts of studies. Nonetheless, to plan and implement a literature study efficiently, Staples and Niazi \cite{Staples:2007:EUS:1282869.1282969} make clear that narrowly defined research questions are key. We therefore recommend to use a combination of generic research questions (e.g., Table~\ref{tab:StandardResearchQuestions} to ``get a feeling'' about the result set) and specific narrow research questions---even for mapping studies.}.

To direct the study towards its goal, i.e., a mapping study or a literature review, further standard questions can be asked that support the next steps in the study selection process. For instance, the scientific maturity addresses the classification according to the \emph{research type facet} \cite{Wieringa:2005:REP:1107677.1107683} to work out the level of evidence in the publications. A mature field should for example not only contain solution proposals, but also validation and evaluation research papers, and consequently experience reports (Figure~\ref{fig:PublicationFrequencySample}). The question for the result set's contribution aims at working out the different kinds of \emph{contribution type facets} \cite{PFMM08} and their respective distribution in the publication population. For instance, does the result set contain models, theories, lessons learned, or frameworks? 
The remaining questions address further general aspects, such as observable streams in the result set. Such streams can become obvious by certain trends or accumulations of publications, e.g., outstanding number of solution proposals and, at the same time, no theories. Such a discussion can also be supported by applying further specific models, such as the \emph{rigor-relevance model} proposed by Ivarsson and Gorschek \cite{Ivarsson:2011:MER:1969653.1969705}. Mainstreams can also be brought to light by studying the contents of the paper in more detail, e.g., by introducing \emph{focus type facets} \cite{Paternoster20141200}, which can also direct the in-depth investigation of a systematic review.

In summary, the standard research questions from Table~\ref{tab:StandardResearchQuestions} aim at providing a demographic overview of the study. Answering these questions shows how many publications have been published over time, about which topics they are, and which results they provide. These questions already provide a big picture of a research field, and they allow for getting a better understanding of the studies available in that field. Finally, these questions also help scoping the study and preparing the collection and selection procedures according to the overall study objectives. For example, an initial analysis of the demographic information helps checking the suitability of research questions and adjusting them if necessary.

\subsubsection{Search Strings}
\label{sec:Approach:Preparation:SearchStrings}
Once the scope of the study has been set, researchers need to reflect on proper search strings, which also depend on the domain under investigation. Depending on the precision of the search strings, the queries may produce inappropriate results, too much overhead, or just an incomplete result set. Therefore, search strings must be defined with care \cite{Kitchenham:2015rt}, and search queries should always be tested prior to the actual search\footnote{Note that the construction of search strings also depends on the planned search strategy (see Sect.~\ref{sec:Approach:DataCollectionCleaning}), since search stings for automated database searches have a different ``layout'' than those used for a curiosity-driven or trail-and-error search, e.g., using Google Scholar. Regardless of the search strategy, finding the proper key words is crucial. The most straight-forward approach to develop appropriate search strings is either to do a trail-and-error search or to call in domain experts. Alternatively, a preliminary study can be conducted to ``test'' the field of interest.}. There exist some strategies to develop proper search strings, e.g.:

\paragraph{Snowballing}
One way to narrow down the search space in advance is to conduct a preliminary investigation of the field by relying on \emph{snowballing} \cite{Kitchenham:2015rt}. That is, the investigation starts by studying publications known in advance and by iteratively extending the known literature set by following the references provided therein. This procedure helps providing an initial overview of the publication space and key contributors, but very much depends on the expertise necessary to select an appropriate starting point (see also Sect.~\ref{sec:ExamplesAndLessons}). However, as reported by Badampudi et al.\ \cite{Badampudi:2015:EUS:2745802.2745818}, manual search strategies compared to automatic ones are capable of producing ``competitive'' results regarding result set precision while, at the same time, avoiding vast overhead usually produced by automatic database searches.

\paragraph{Trail-and-Error Search}
One approach suitable to find and test search queries is the ``Trail-and-Error Search''. This approach relies on meta-search engines, e.g., Scopus or Google Scholar, and requires initial keywords or (partial) key phrases that are considered search query candidates for the ``real'' search. The purpose aims at iteratively narrowing down the list of potential candidates by checking whether:
\begin{itemize}
	\item A search query returns a (potentially) meaningful result set.
	\item A keyword or a combination thereof returns hits (at all).
	\item A search query is of sufficient precision; for instance, if searching a particular domain, how many hits are not in the domain of interest?
\end{itemize}
Hence, a trail-and-error search serves two major purposes: First, it can be used to initially test and develop search queries, e.g., by determining which keywords might (not) generate useful results. Second, results from such test runs can be used to harvest reference publications to support manual search strategies (as for instance exercised in \cite{Theocharis:2015aa}). Although this approach can be seen as everything but a good scientific practice, it still helps taking the initial steps into the overall research design development---especially in domains in which few or no secondary studies are present to provide structure to the field of interest (as it for instance was the case in \cite{Ingibergsson:2015aa}).

\subsubsection{Inclusion and Exclusion Criteria}
\label{sec:Approach:Preparation:ICEC}
Depending on the study's scope, result sets can contain a vast amount of potentially relevant publications. In the worst case, we experienced searches to yield in several thousands of hits. We doubt it should be questionable that several 10,000 hits cannot be treated seriously within an acceptable timeframe\footnote{As it is also criticized by Staples and Niazi \cite{Staples:2007:EUS:1282869.1282969}. In \cite{Kuhrmann:2015ix}, however, we accepted this challenge. It took us about a year just to clean the data and perform the selection procedures. We do not recommend this for replication.}. Therefore, researchers need to clean the dataset and to select the relevant studies. In order to make these procedures rigorous and reproducible, inclusion and exclusion criteria need to be defined.
\begin{table*}[!ht]
\centering
\caption{Exemplary (standard) inclusion (I) and exclusion (E) criteria for literature studies.}
\label{tab:StandardInExclusionCriteria}
\begin{tabular}{@{}lc@{\quad}p{0.87\textwidth}@{}}
    \hline\noalign{\smallskip}
    	\textbf{No.} &  & \textbf{Criterion}  \\
    	\noalign{\smallskip}\hline\hline\noalign{\smallskip}
	1 & I & Title, keyword list, and abstract make explicit that the paper is related to [topic]. \\
	2 & I & The paper presents [topic]-related contributions, e.g., [topic list]. \\ 
	\noalign{\smallskip}\hline\noalign{\smallskip}
	3 & E & The paper is not in English [or any other language of interest]. \\
	4 & E & The paper is not in the domain [domain name(s)]. \\
	5 & E & The paper is a tutorial-, workshop-, or poster summary only. \\
	6 & E & The paper relates to [topic] in its related work only. \\
	7 & E & The paper occurs multiple times in the result set. \\
	8 & E & The paper's full text is not available for download. \\
\noalign{\smallskip}\hline\noalign{\smallskip}
\end{tabular}
\end{table*}

Similar as with standard research questions (Table~\ref{tab:StandardResearchQuestions}), we experienced some inclusion and exclusion criteria to be useful in a broad spectrum of studies. These standard criteria listed in Table~\ref{tab:StandardInExclusionCriteria} allow researchers to obtain an appropriate result set and to define their requirements on the objective-dependent relevance of publications retrieved. For instance, experience shows workshop- or tutorial summaries can contain a lot of relevant keywords, but might not necessarily advance the actual body of knowledge. Also, since contributions might occur multiple times or might be out of scope, those have to be eliminated as soon as possible (criterion 7). Another important criterion is the eighth, i.e., if the full text is not available, the respective publication is usually of little value (regarding possibilities to analyze them and eventually draw proper conclusions). In context of a mapping study, this issue can be compensated to a certain extent as those studies focus on an early, abstract-based analysis. However, when it comes to in-depth analyses, e.g., in a systematic review, the full text needs to be available. 

Finally, Kitchenham et al.\ \cite{Kitchenham:2015rt} recommend aligning search strings with the research questions. We add to this the suggestion to also align the in-/exclusion criteria with the research questions. This might result in a number of ``duplicated'' criteria, i.e., a paper could be relevant to topic \emph{A} or to topic \emph{B} if the literature study aims at synthesizing knowledge thus requiring multiple topics to be addressed and analyzed together. This furthermore allows for later replication why a specific paper was in- or excluded to/from the study.

\subsection{Data Collection and Dataset Cleaning}
\label{sec:Approach:DataCollectionCleaning}
Once the study is designed, data can be collected. In that stage, the resulting data needs to be analyzed, cleaned/harmonized, and prepared for the upcoming investigations.

\subsubsection{Data Collection}
\label{sec:Approach:DataCollectionCleaning:Collection}
The data collection is usually conducted as an automated search using different sources. Automated data search, however, needs careful preparation and potentially extra test runs, as every data source has a slightly different format of the query strings, or constraints regarding the queries' length and complexity (see also the discussion in \cite{Ali:2014:ESS:2652524.2652557,Badampudi:2015:EUS:2745802.2745818,Brereton:2007:LAS:1225950.1226109,Kitchenham:2015rt}). 
In practice, we experienced the design of multiple and overlapping query strings beneficial. Although the search procedure must be executed several times and produces some overhead, simple queries are usually better accepted by the search engines (see Sect.~\ref{sec:ExamplesAndLessons} for a detailed discussion).

\paragraph{Appropriate Data Sources}
Depending on the particular disciplines, several standard databases or collections (so-called baskets\footnote{Such as the Senior Scholars Basket, cf.\  \url{http://home.aisnet.org/displaycommon.cfm?an=1&subarticlenbr=346}}) are available. In the following, we give an exemplary discussion for software engineering. Apart from specific conference- and workshop series (so-called restricted approach \cite{Kitchenham:2015rt}), a literature search should address the most common sources. That is, instead of searching specific proceedings of a conference, search queries should be designed to work with entire digital libraries. For the more general field of software engineering, the following libraries can be considered as standard libraries (or subsets thereof if opting for the restricted approach):
\begin{itemize}
	\item IEEE Digital Library (Xplore)
	\item ACM Digital Library
	\item SpringerLink
	\item ScienceDirect (Elsevier)
	\item Wiley Interscience
	\item IET (also accessible via IEEE)
\end{itemize}
However, these libraries have their ``specialties'', notably, regarding the search query construction. Another point to take care of when using such digital libraries is the continuous indexing, i.e., indexes will ``evolve'' over time, which makes it hard to reproduce searches (see Sect.~\ref{sec:LL:SearchStringsAndEngines}).

\paragraph{Checking the Result Set}
Before conducting the data collection, we recommend to have a set of reference publications available. One criterion we found useful for checking the appropriateness of a search is if the result set contains the expected reference publications (see also, e.g.,~\cite{Zhang:2011:IRS:1968229.1968314}). If one expects a particular publication in the result set, e.g., arising from a preliminary search, but it is not contained in that set, the revision of the search might be recommendable. Options to identify reference publications can be found in Sect.~\ref{sec:Approach:Preparation:SearchStrings}.

\paragraph{Primary Search and Backup Search}
Primary searches should always be conducted using aforementioned (or comparable) standard libraries. However, for several reasons, those libraries do not always contain all relevant publications. For example, contributions relevant to the field might result from Ph.D.\ theses that are not published in/not indexed by the standard libraries. 

Therefore, we experienced it beneficial to complement the primary search with a backup search utilizing meta-search engines to complete the result set. However, using a meta-search engine must be done carefully. Besides the standard meta-search engines\footnote{Note: Apart from serving the backup search, meta-search engines can also be a useful instrument in studies that also include (continuous) updates, e.g., to monitor the development of a field over time \cite{Kuhrmann:2016gf}.}, such as DBLP or Scopus, Google Scholar is often used to get results quickly. However, the quality of search results obtained from such engines also depends on search preferences and even trends and, thus, searches might be much less repeatable than compared to standard libraries. Also, the results might also provoke duplicates and introduce extra threats to the validity of a literature study. A Ph.D.\ thesis, for example, can be written in a cumulative manner where parts of it exist separately as peer-reviewed publications already present in the result set of the primary search.
Hence, it is important that the results obtained via meta-search engines are not included into the main result set without crosscheck. To this end, hits produced by such engines should be included in an own category, and such searches should be discussed as part of the threats to validity to increase the transparency.

\paragraph{(Data) Export Practices}
Data obtained from a data source must be stored in a way in which it can be used for further analyses. This part can become time consuming since different databases provide different export formats, which later on need to be joined and integrated. Therefore, data should be exported in at least two formats:
\begin{itemize}
	\item A literature management tool of choice, such as \BibTeX 
	\item As plain or (better) comma-separated (CSV) text files
\end{itemize}
These formats have the advantage that they are easy to process and convert into spreadsheets to allow for further selection (Sect.~\ref{sec:Approach:DataCollectionCleaning:Cleaning}), and later on, analysis steps.

\subsubsection{Dataset Cleaning}
\label{sec:Approach:DataCollectionCleaning:Cleaning}
Cleaning a result set is a demanding, time-consuming task. Usually, we find two types of papers to be removed from the result set (cf.\ Table~\ref{tab:StandardInExclusionCriteria}):
\begin{enumerate}
	\item Contributions that are out of scope, and 
	\item Duplicates.
\end{enumerate}
Duplicates are easy to find and eliminate, yet it is hard to decide which of the duplicates should be eliminated. It often happens that one publication is listed in multiple literature databases (e.g., for cross-indexing reasons). In such cases, it needs to be decided which paper to consider for inclusion into the result set. A pragmatic approach is to include the results from the database that provides the paper for download and to remove the other occurrences; this needs, however, to be defined in the exclusion criteria for the sake of transparency. Another case for a duplicate is a conference paper, followed by a journal article, e.g., a special issue paper. In such cases, it must be decided whether the original or the extended publication should be selected for inclusion. A criterion could be to always select the higher-valued publication, i.e., journal over conference, as journal articles are expected to have a higher maturity \cite{Paternoster20141200} and level of detail.

Publications that are out of scope are, on the other hand, easy to remove, yet they are often hard to identify if part of a large result set. Since the result set might have been created from an automatic search, even out-of-scope publications that met at least one selection criterion could be present. Those publications need to be found manually and removed in the cleaning procedures. 

\paragraph{Scoping via Word Clouds}
To support the identification of out-of-scope-papers, we experienced word clouds (tag clouds) to be a useful tool. Word clouds can be automatically created using keyword lists or abstracts. A word cloud is an instrument to visualize the (quantified) occurrence of a word/term in relation to other terms. They can be easily created using several publicly available tools\footnote{Note: Some of the tools have limitations regarding the amount of text they can process. Furthermore, the tools offer different features, such as thresholds, visualization and export mechanisms. Those points need to be evaluated prior to usage.}, e.g., Wordl or TagCrowd\footnote{Both tools are available at: \url{http://www.wordle.net} and \url{http://tagcrowd.com/}}.
\begin{figure*}[!t]
  \includegraphics[width=\textwidth, right]{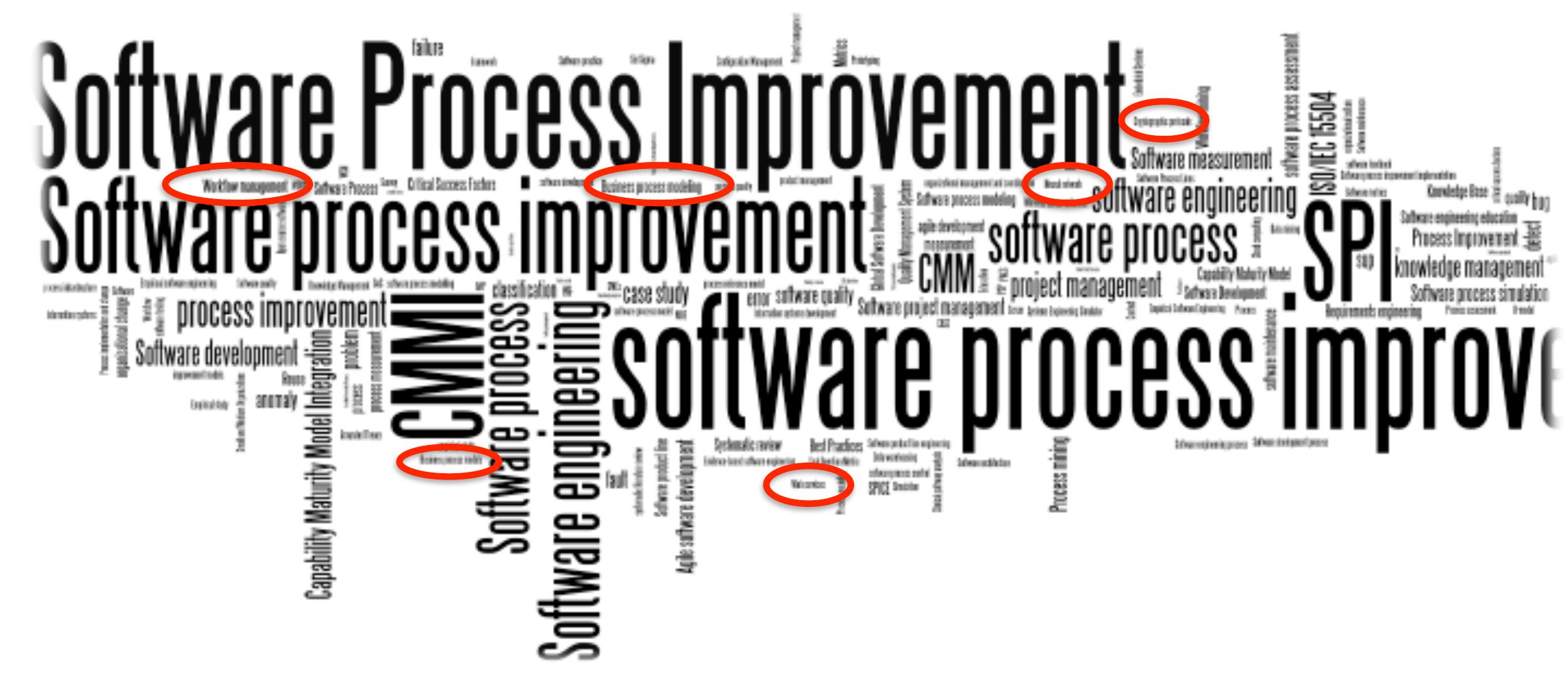}
  \caption{Example of a word cloud from \cite{Kuhrmann:2015ix} for visually inspecting the result set. ``Outliers'' to be used for excluding further papers from the result set are highlighted.}
  \label{fig:WordcloudSample}
\end{figure*}

Word clouds can serve two purposes: First, word clouds can be used to analyze the appropriateness of a result set. A word cloud, which is based on the keywords, can be analyzed to work out whether the contained publications' keywords match the expectations (Figure~\ref{fig:WordcloudSample}). Unexpected and/or ``wrong'' keywords can be easily detected and used to clean the result set. Depending on the quality, a considerable share of non-fitting papers can be removed; remaining papers (in a reduced set) are then removed during the selection phase (Sect.~\ref{sec:Approach:StudySelection}). 

However, word clouds have to be used with care: even though there is research that shows word clouds providing improvement concerning the clustering and summarizing of descriptive information, such as \cite{Oosterman:2010:ECT:1952222.1952284,Ramage:2010aa,Kuo:2007:TCS:1242572.1242766,Schrammel:2009:SST:1518701.1519010,Rivadeneira:2007:GOH:1240624.1240775}, there is still the risk of eliminating relevant papers; for instance, because those papers might rely on a rarely used terminology. Therefore, eliminating papers based on word clouds only might threaten the validity of a study why we recommend that the use of word clouds must be planned with care and in detail in advance, and resulting candidates for removal require careful inspection.

As a second purpose to be served, a word cloud can support the later analysis of a result set during, for example, the concept classification conducted as part of a mapping study. For instance, in our study on method engineering \cite{Kuhrmann:2014vn}, we analyzed the final word cloud to get a better understanding about which research type facets to expect from the publication set (e.g., how to interpret terms like ``case study'' as used in the respective community). The result of the word cloud and the result of the classification conducted in the study can then be compared to analyze the subjective authors' self-classification and the more objective one from the reviewers' classification. In another example \cite{Kuhrmann:2016ul}, we used a word cloud to support the development of a \emph{focus type facet} \cite{Paternoster20141200} and, furthermore, to conduct a cluster analysis. 

\begin{figure*}[t]
  \sidecaption
  \includegraphics[width=0.67\textwidth]{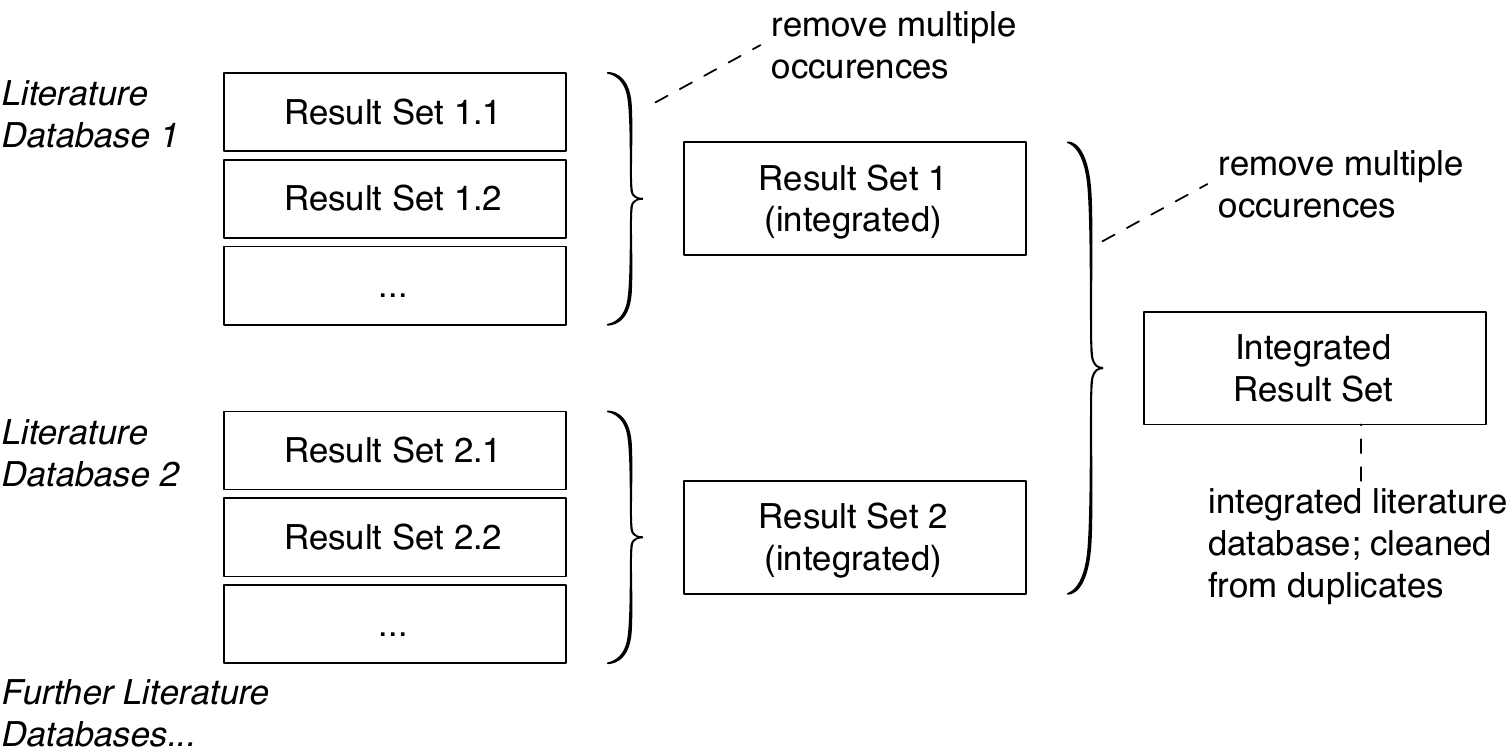}
  \caption{Exemplary procedure of stepwise integrating and cleaning literature databases. In each integration step, reporting-relevant information needs to be recorded.}
  \label{fig:04:DataCleaningProcedure}
\end{figure*}

\paragraph{Merging and Reducing the Integrated Dataset}
Depending on the particular search strategy---especially the search query construction approach---researchers have to deal with multiple (isolated) datasets. This is especially true if the work during the data collection is distributed among multiple researchers. To prepare the selection, the individual result sets need to be integrated into a holistic one. This integration constitutes a challenging task:
\begin{itemize}
	\item If a literature database was queried multiple times (e.g., for the search string construction), the individual results need to be joined.
	\item Every literature database provides a slightly different export format and/or structure, e.g., CSV files obtained from Springer and from ACM have a different structure. These differences need to be reconciled.
	\item If duplicates were removed on a per-database basis, the integrated result set may still contain cross-database duplicates. The integrated dataset must then be cleaned again by identifying and removing duplicates.
	\item If the individual datasets were yet not investigated for duplicates, the respective cleaning procedures must be performed now.
\end{itemize}
The aforementioned steps can be (partially) automated \cite{Kuhrmann:2016gf}. Nevertheless, the inclusion and exclusion criteria selected for the study should be consulted to support the compilation of the integrated dataset as well. We experienced the following procedure (Figure~\ref{fig:04:DataCleaningProcedure}) to be best suited for the stepwise integration: 
\begin{enumerate}
	\item Integrate and clean the data on a per-database level, i.e., if a database was queried multiple times, integrate the obtained sub-result-sets first.
	\item Integrate all sub-result-sets into the integrated dataset and repeat the cleaning. 
\end{enumerate}
Eventually, we create an integrated dataset. Appendix~\ref{sec:app:DataStructures} provides an example illustrating and explaining the minimal required data. Please note that the step of integrating and reducing the data is crucial and, therefore, needs to be documented carefully. The particular steps of the applied procedures are valuable information for other researchers to reproduce the overall study. Furthermore, the outcome of these steps forms the input for the rest of the study. Hence, researchers must ensure that no relevant publication is lost during this step.

\paragraph{Step-wise Dataset Completion}
Once an integrated data set is obtained, it should be analyzed for (sufficient) completeness. Depending on the database and the individual publications, some information might be missing, e.g., abstracts or keywords. This information needs to be collected and integrated, however, the step bears some pitfalls:
\begin{itemize}
	\item There are abstract-free publications, e.g., magazine articles, of which the respective literature databases provide parts of the introduction section as abstract substitute. Such cases require a manual inspection and researchers need to discuss how to treat them.
	\item There are publications without (electronically available) keywords. These are publications that have no keywords at all, or publications that may well have defined keywords, but those were not listed in the exported data structure. For those publications, it must be defined how to treat them.
	\item For technical reasons, some literature databases do not provide options to export the abstracts. In such cases, manual work is required to get the abstract and integrate it into the dataset.
	\item Pieces of required metadata might be missing, e.g., the publication year, publication vehicle (conference, journal, etc.). This information needs to be completed.
\end{itemize}
Apart from this essential information, another aspect needs to be taken into account: the representation of the authors. Literature databases do not have a uniform representation of the author lists; for instance, authors might have varying affiliations or their first and second names are ordered differently (e.g., ``J. J. Abrams'' versus ``Abrams, J. J.''). If researchers plan for a study, for example, to conduct some analyses on the author lists, such as by creating collaboration networks, cliques, and mainstreams, the author information must be available in a uniform way. As dataset completion can be extremely fidgety work, it should be performed iteratively and under continuous quality assurance:
\begin{enumerate}
	\item Complete the abstracts
	\item Complete the keywords
	\item Complete all other required metadata
	\item Ensure consistency in the author lists
\end{enumerate}

\paragraph{Dataset Structure: A Template}
To support all aforementioned steps, a defined data structure needs to be in place. The particular data structure depends on the specific study. However, we recommend minimal data structure shown in Table~\ref{tab:StandardDataCollectionDataStructure} (Appendix~\ref{sec:app:DataStructures}) as it emerges from our previously conducted studies. The table illustrates the recommended minimal data structure to organize the result set. This table serves the basic purposes and can be extended respecting the actual study's needs, such as extra columns for classifications for mapping studies.

\subsection{Study Selection}
\label{sec:Approach:StudySelection}
In the study selection phase, the prepared dataset is analyzed for publications relevant for the actual study, i.e., researchers systematically select the relevant papers from the search results (this phase relates to the (primary) study selection in \cite{Kitchenham:2015rt}). Since result sets can comprise several hundreds or even thousands of papers, this phase requires special attention and, thus, careful planning.

\subsubsection{Plan: Defining the Study Selection Procedure}
\label{sec:Approach:StudySelection:Plan}
Many factors influence the actual study selection (e.g., number of researchers, degree of distribution, familiarity with the topic, etc.). In case of multiple researchers conducting the study, we consider the following aspects of the study selection necessary to be planned and agreed on in advance:
\begin{itemize}
	\item Schedule for the study selection including workshops, regular meetings/calls for discussing intermediate selection/classification results, etc.
	\item Technical infrastructure (tools, data storage, file formats, etc.)
	\item The criteria upon which researchers decide the relevance of a publication
	\item The procedure to infer an agreement and the voting procedure (if applicable)
\end{itemize}
The last step, the voting, assumes that various researchers vote for in-/exclusion of a publication independently. The final decision for including the publication into the final result set then depends on the result of the voting. There are many practices that can be included into the voting procedure (e.g., veto rights) while we believe that this also much depends on the research context, e.g., researchers' experience, expertise, but also their personal preferences to conduct the study (see also Sect.~\ref{sec:Approach:StudySelection:VotingProcedure}). 

\subsubsection{Kick-Off: Setting Up the Selection Approach}
\label{sec:Approach:StudySelection:KickOff}
Assuming a study within a group of multiple researchers, the study selection starts with a kick-off meeting in which the inclusion and exclusion criteria are recalled (Table~\ref{tab:StandardInExclusionCriteria}), the selection/voting procedure is discussed, and a schedule for subsequent meetings is defined. In the following, every participating reviewer gets a copy of the cleaned result set, which is rated individually. That is, the study selection procedures are initiated.

\subsubsection{Voting Procedure}
\label{sec:Approach:StudySelection:VotingProcedure}
Voting is essentially a headcount procedure in which a team of researchers works out a decision whether a particular paper is considered relevant for the study or not, i.e., to eliminate those papers from the result set that are considered irrelevant. The relevance can be determined by different measures, which need to be defined in advance (e.g., title, abstract, and full text). Potential routes towards a decision are \emph{majority votes} or \emph{relative ratings}. The actual classification can be carried out in a group of researchers or individually, iteratively, round-based or in workshops. In the following, we focus on an individual, traditional round-based classification. 

\begin{figure*}[!ht]
  \sidecaption
  \includegraphics[width=0.67\textwidth]{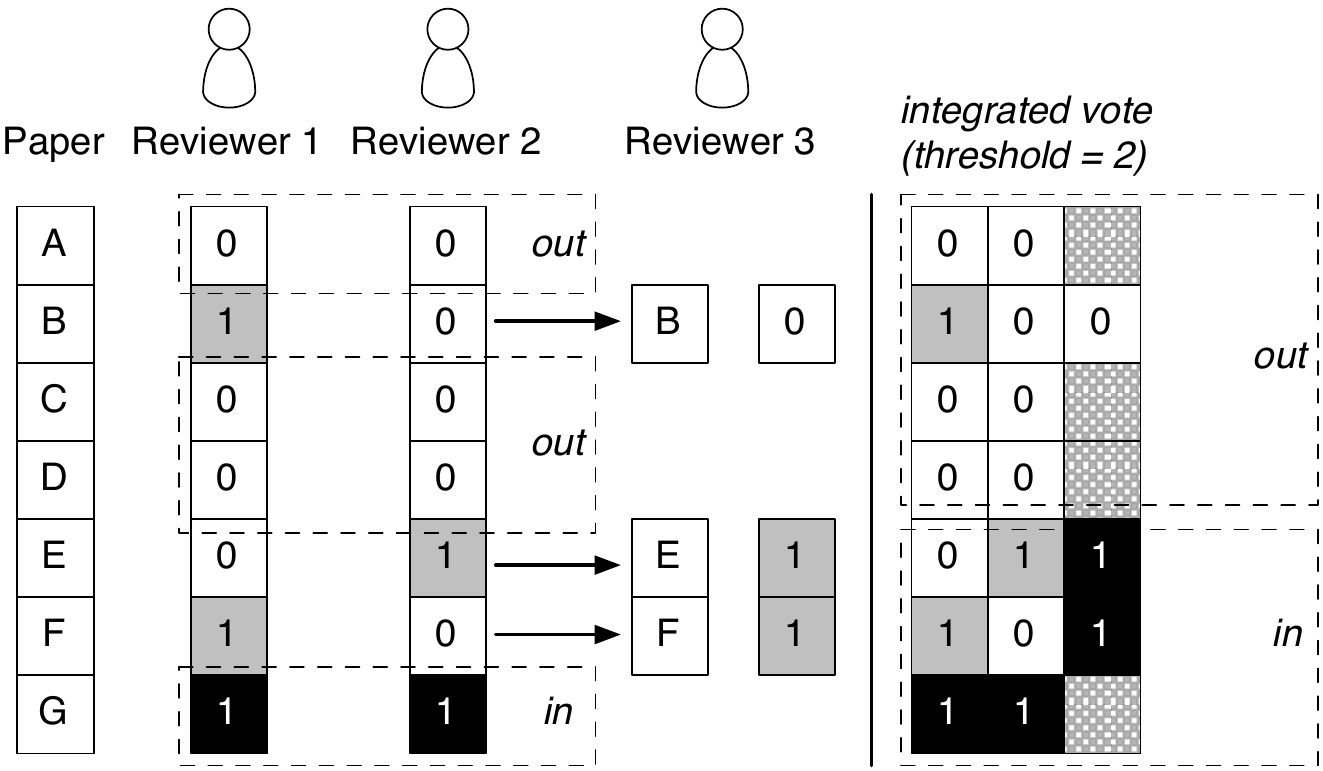}
  \caption{Overview of the standard majority voting procedure for a 3-reviewer team.}
  \label{fig:05:StandardMajorityVoting}
\end{figure*}

\paragraph{Majority Voting}
The voting is a headcount that aims to bring in objectivity into the study selection. Although there are in-/exclusion criteria, the final application of the criteria to the publications to be selected is in the hands of individuals thus including individual interpretations of a publication. The reason why we recommend including multiple researchers in this procedure is to overcome the inherent threat arising from this subjectivity. Hence, we also consider a \emph{majority vote} to be the standard procedure as it is the most straightforward approach: every reviewer is provided with the integrated result set and reviews the items individually according to defined criteria, e.g., title and/or abstract. If a reviewer considers a publication relevant, 1 point is given, 0 otherwise. For \emph{n} reviewers and \emph{m} publications, the procedure results in an $n\times m$ voting matrix, which helps to select the relevant papers.
The (final) selection is then based upon the agreements, such as a threshold or agreement statistics (e.g., Cohen's or Fleiss' $\kappa$). For example, if three reviewers participate, the voting procedure could be organized as shown in Figure~\ref{fig:05:StandardMajorityVoting}: two reviewers start individually. To get a paper included in the set of relevant papers, 2 points are required (threshold approach). Two reviewers can come up with the following results: 2 points = paper is relevant, 0 points = paper is irrelevant, and 1 point = paper is not yet decided. In the next step, the third reviewer\footnote{Please note that a reviewer can be an internal reviewer (e.g., a co-author) as well as an external researcher or expert not involved at all in the design in case of unknown domains.} is called in and is presented a reduced list that only contains the papers yet not decided. The third reviewer then conducts the voting to finally decide about the papers' relevance.

\begin{figure*}[!t]
  \sidecaption
  \includegraphics[width=0.67\textwidth]{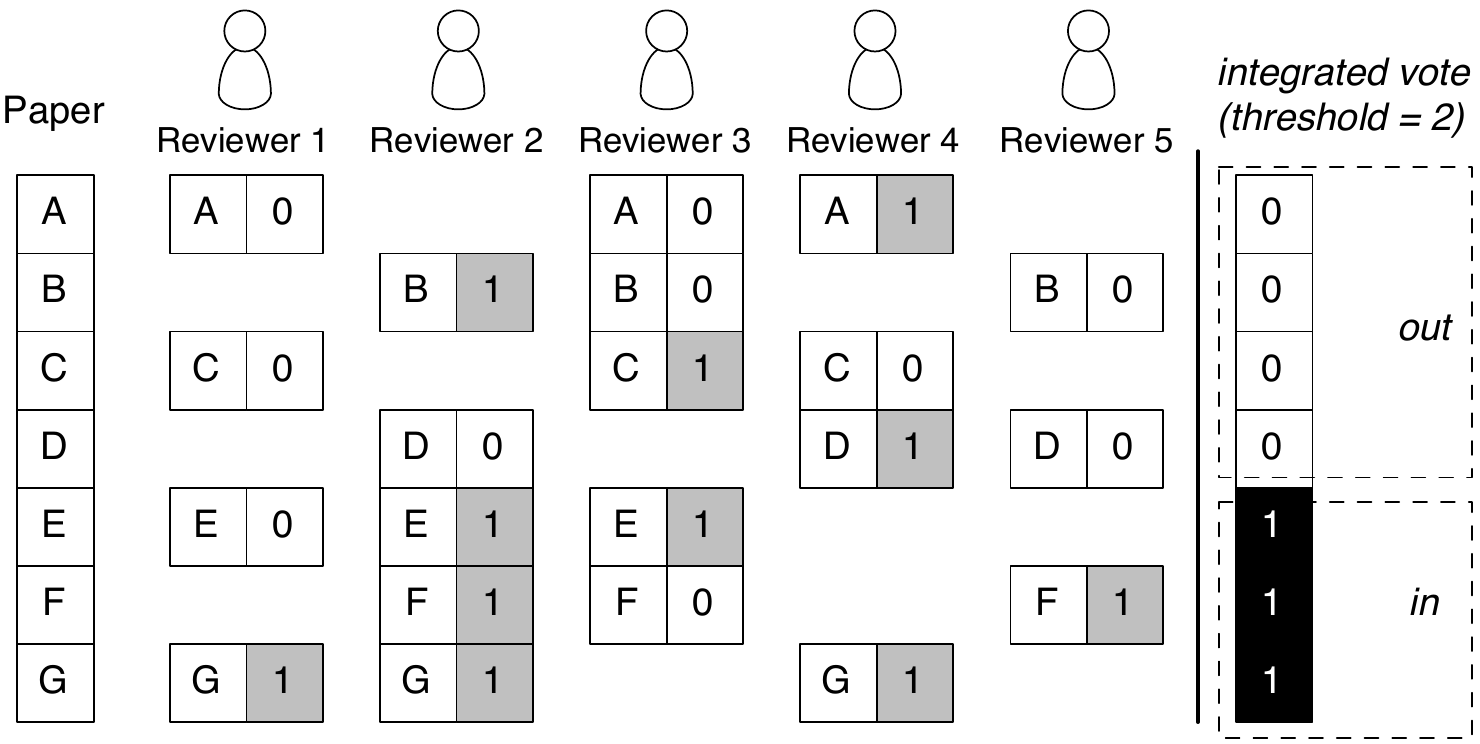}
  \caption{Paper selection based on overlapping paper subsets (a reviewer evaluates only subset of papers, usually just one run required to find the selection).}
  \label{fig:06:OverlappingSubsets}
\end{figure*}

\paragraph{Alternative Approaches}
Instead of calling in a third reviewer to conduct a fully independent review, a voting workshop can be organized. In such a workshop, all reviewers involved in the selection process discuss and decide the non-decided papers. We applied this approach for instance in \cite{MOWD14,Kuhrmann:2015ix}. 
Yet another approach is to provide reviewers with overlapping subsets of the whole result set, e.g., to incrementally collect three votes in just one run (Figure~\ref{fig:06:OverlappingSubsets}).

\paragraph{Scaling} 
So far, we performed the majority voting procedure with 2 reviewers in the workshop model, 3 and 4 reviewers, and two 2-person review teams (see Sect.~\ref{sec:ExamplesAndLessons}). However, as we talk about simply summing up points, the approach can be scaled to an even larger number of reviewers. A paper's relevance is then simply defined by a function
\begin{equation}
\label{math:BasicRelevance}
	\emph{relevance}:\mathbb{R}^{+}\!\times\ \mathbb{Z}\to \{0, 1, ?\}
\end{equation}
that is used to determine the relevance of a paper $p_{j}$ in relation to a threshold \emph{th}, and to (de-)select papers or marking them for later decision:
\begin{equation}
\label{math:RelevanceDetermination}
\emph{relevance}(\emph{rating}(p_{j}), \text{th}) =
  \begin{cases}
    1	& \quad \text{if } \emph{rating}(p_{j}) > \text{th} \\
    0	& \quad \text{if } \emph{rating}(p_{j}) < \text{th} \\
    \text{toDecide}	& \quad \text{if } \emph{rating}(p_{j}) = \text{th} \\
  \end{cases}
\end{equation}
The actual threshold \emph{th} needs to be defined during the initialization of the selection procedure (Sect.~\ref{sec:Approach:StudySelection:Plan}). The  rating  (simple, unweighted case; Figure~\ref{fig:05:StandardMajorityVoting}) of a paper is then defined by the number of points that a paper received from $n$ reviewers involved in the process:
\begin{equation}
\label{math:Rating:HeadCount}
	\emph{rating}(p_{j})=\sum^{n}_{i=1}r_{i}(p_{j})
\end{equation}
Regardless of the number of stages and reviewers involved, rating statistics need to be carefully documented in order to be able to reproduce which paper came in in which stage and to make explicit the inter-rater agreement. Furthermore, we also suggest to document according to which criteria a paper was included or excluded after all, which can require extending the data structure of the result set to keep this information.

\paragraph{Relative Rating}
The \emph{relative ratings} approach\footnote{So far, we did not yet apply this method to a complete study, but partially applied it during sample-based result set testing and evaluation (cf. Sect.~\ref{sec:ExamplesAndLessons}). As this approach is quite complex compared to the majority vote, it requires sufficient tool support.} as illustrated in Figure~\ref{fig:07:RealitveVoting} is similar to the \emph{majority voting} where all reviewers are asked to vote a result set, but with a difference in the applied metric: Instead of a simple ``Yes/No'' (1/0) metric, in this approach, we use Likert scales and thresholds. The basic underlying procedure remains the same: each reviewer is provided with the integrated result set and rates a paper, but on a scale, such as a 5-point Likert scale:
\begin{figure*}[t]
  \sidecaption
  \includegraphics[width=0.67\textwidth]{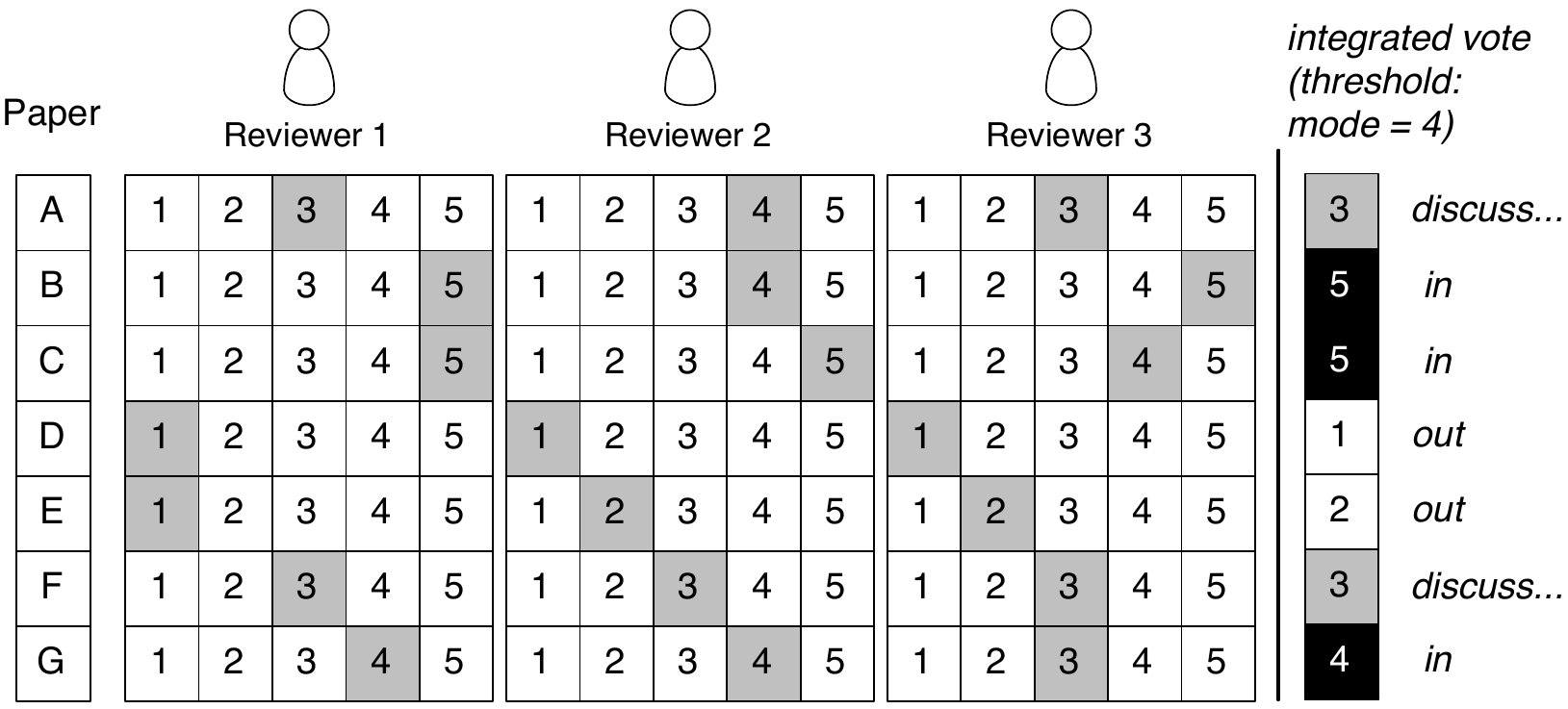}
  \caption{Paper selection based on relative votes (final selection is, in this example, made using the \emph{mode} value, while the ``neutral'' element 3 serves as marker for papers to be discussed).}
  \label{fig:07:RealitveVoting}
\end{figure*}
\begin{itemize}
	\item 5 points: Paper is highly relevant (must be included)
	\item 4 points: Paper is (somewhat) relevant
	\item 3 points: neutral/no opinion
	\item 2 points: Paper is not relevant
	\item 1 point: Paper is absolutely irrelevant
\end{itemize}
Based on the individual ratings, relevance can be determined, e.g., using the mean value or the mode, and precision can be determined, e.g., using standard deviations or distance metrics. The inclusion criterion is then a selected value on the used scale, e.g., 4 (or better). Critical is the handling of papers that end up with the neutral value. These papers require extra handling.

\paragraph{Balancing Votes}
How reliable is this way of selecting papers? In the simple case, which is the \emph{majority vote}, a democratic headcount is used to in-/exclude a paper. However, this procedure has some flaws. For instance, given a situation in which two reviewers ended up in a stalemate. A third reviewer is then called in to make the decision; and now scale this up to 7 reviewers: the 7th reviewer makes the final decision by outvoting 3 others. To overcome such situations, workshops can be performed to discuss critical papers (which can be unrealistic if, for instance, 250 papers need to be discussed), thresholds can be defined, or weights can be introduced, e.g., senior reviewer votes count twice.
However, the basic problem remains: what is the level of agreement, i.e., the reliability of the selection? As a first step to determine the reliability, the inter-rater reliability can be calculated, e.g., using Cohen's $\kappa$ \cite{Cohen:1968aa} for two reviewers or, more general, Fleiss' $\kappa$ \cite{Fleiss:1971aa} for more than two reviewers\footnote{Please note that inter-rater reliability calculations also depend on the scales applied, e.g., weighted $\kappa$ values when using ordinal data (cf.\ \cite{Kitchenham:2015rt,Wohlin-C.:2012uq}).}. Furthermore, the basic agreement can be visualized (and partially automated) as shown in Figure~\ref{fig:ColorCodedSpreadsheet} (Appendix~\ref{sec:app:DataStructures}).

Yet, the headcount is a fairly simple, but absolute metric. In some cases, we experienced the need for a more differentiated vote, which can be implemented, e.g., using \emph{relative votes} with Likert scales. However, the more differentiating scale introduces a new challenge: \emph{How to find a final and consolidated rating?} Approaching a consolidated rating via the mean or the mode might fail, because they are easy to trick or because they might be even not applicable; consider, for example, the mode of $\{0,0,1,1\}$, and what a resulting mean of 0.5 even implies in relation to a $\emph{th}\in\mathbb{Z}$ (cf.\ Eq.~\ref{math:BasicRelevance}). Again, a simple solution could be to introduce rater-specific weights. Furthermore, simple weighting methods, such as, the 3-point-method can be applied, with $V_{j}=\{v_{p_{j}}^{r_{1}},...,v_{p_{j}}^{r_{n}}\}$ being the set of $n$ reviewer votes for a paper $p_{j}$: 
\begin{equation}
\label{math:WeightedRating}
	\emph{rating}(p_{j})=\left|\frac{min(V_{j})+4\cdot \bar V_{j} + max(V_{j})}{6}\right|
\end{equation}
The extended weighted rating from Eq.~\ref{math:WeightedRating} can be used in the determination of the relevance in Eq.~\ref{math:RelevanceDetermination}.

\subsubsection{The Gathering: Integrate and Finalize the Paper Selection}
\label{sec:Approach:StudySelection:TheGathering}
Having all individual ratings conducted, the study's moderator (Kitchenham et al.\ \cite{Kitchenham:2015rt} speak of a team leader) collects all individual ratings and starts the integration of the results.
The basic task is to, initially, integrate the individual ratings to work out the current state of selected and/or undecided papers (see also color-coding in Figure~\ref{fig:ColorCodedSpreadsheet} that is based on Eq.~\ref{math:RelevanceDetermination} and Eq.~\ref{math:Rating:HeadCount}). Depending on the approach defined in the initialization of the selection procedure (Sect.~\ref{sec:Approach:StudySelection:Plan}), the moderator prepares the dataset for extra review iterations and/or organizes required workshops. In the following, the selection procedure is iterated until all papers are finally decided. 

Once all papers are decided, the moderator draws a baseline and prepares the final selection of papers, i.e., a cleaned list that only contains those papers considered relevant for the study, and he finally prepares the clearing work.

\subsubsection{Class Dismissed: Analyze the Result Set and Report}
\label{sec:Approach:StudySelection:ClassDismissed}
When the selection is done, the moderator concludes the selection process and prepares the handover to the actual analysis. This includes some standard tasks as well as some optional tasks depending on the eventually targeted study. In particular, the moderator has to prepare the study selection report and the resulting literature database. The literature database must at least contain all papers that were selected as relevant to the study. The report comprises some statistics, such as, databases, results per database from search, and elimination statistics (an example is shown in Table~\ref{tab:SurveyResults}).
\begin{table*}[!t] 
\renewcommand{\arraystretch}{1.2}
\caption{Exemplary search and selection report (excerpt from \cite{Kuhrmann:2015ix}).}
\label{tab:SurveyResults}
\begin{center} 
	\begin{tabular}{p{0.45\textwidth}rrrr} 
		\hline\noalign{\smallskip}
		\textbf{Step} & \textbf{IEEE} & \textbf{ACM} & \textbf{\ldots} & \textbf{Total} \\
		\noalign{\smallskip}\hline\hline\noalign{\smallskip}
		\emph{Step 1: Search}      &          &       &       &       \\
		\hline
		$\text{S}_{1}$ \textbf{and} ($\text{C}_{1}$ \textbf{or} $\text{C}_{2}$) &	71    & 543   & \ldots & 3,185 \\
		\ldots       & \ldots & \ldots & \ldots & \ldots \\
		$\text{S}_{8}$ \textbf{and} $\text{C}_{2}$	       & 	114   & 105   & \ldots &   8,374 \\
		\hline
		\emph{Step 2: Removing Duplicates}      &          &       &        &       \\
		\hline
		Duplicates per database & 1,486 & 566 & \ldots & 16,643 \\
		Duplicates across all databases & 916 & 551 & \ldots & \textbf{5,315} \\
		\hline
		\emph{Step 3: In-depth Filtering}      &          &        &     &       \\
		\hline
		Applying filters $\text{F}_{1}$ and $\text{F}_{2}$ & 578 & --  & \ldots & 1,562 \\
		Unfiltered                        & --  & 551 & \ldots & 1,610 \\
		\hline   
		Result set (search process) & 578 & 551  & \ldots & \textbf{3,172} \\
		\hline
		\emph{Step 4: Voting}       &       &        &     &       \\
		\hline
		\textbf{Final result set}      &   283   &  65  &  \ldots  &   \textbf{635} \\
		\noalign{\smallskip}\hline\noalign{\smallskip}
	\end{tabular}
\end{center}
\end{table*}

Depending on the intended study type, just in this step, the moderator can also provide some extra data to support the later analyses. For example, if applicable, the inter-rater agreement helps identify those publications that form the heart of the result set. Furthermore, several outputs can be generated from the result set that help finding a starting point, e.g., exports of the keyword lists and abstracts and word clouds generated thereof, and, associated with more effort, social networks (Sect.~\ref{sec:LL:PreparingTheHandover}). 

\subsection{Concluding and Handover to Data Analysis}
\label{sec:Approach:Concluding}
The last step consists in initiating the actual data analysis, which is dictated by the research questions and eventually the type of secondary study. From the aforementioned described steps, the outcomes listed in Table~\ref{tab:Outcomes} have to be assembled and shipped to the in-depth analyses.
These deliverables can be properly integrated with the research protocols as, for instance, recommended by Kitchenham et al.\ \cite{Kitchenham:2015rt}.

\begin{table*}[ht]
\centering
\caption{Artifacts to be created in the early stages of literature studies to be shipped to the in-depth data analysis.}
\label{tab:Outcomes}
\begin{tabular}{@{}l@{\quad}p{0.85\textwidth}@{}}
    \hline\noalign{\smallskip}
   	\textbf{Reference} & \textbf{Outcomes and content to be delivered}  \\
   	\noalign{\smallskip}\hline\hline\noalign{\smallskip}
	Sect.~\ref{sec:Approach:Preparation:SearchStrings}         & Search terms and resulting search queries (generic terms and queries, as well as database-specific queries) \\
	Sect.~\ref{sec:Approach:Preparation:ICEC}                  & In-/exclusion criteria used in the study \\
	Sect.~\ref{sec:Approach:DataCollectionCleaning:Collection} & List of selected and queried databases, and raw result sets (e.g., CSV files) \\
	Sect.~\ref{sec:Approach:DataCollectionCleaning:Cleaning}   & Cleaned and integrated data sets (including all support instruments used) \\
	Sect.~\ref{sec:Approach:StudySelection:Plan}               & A documented study selection approach, including team setup, selection procedures, and so forth \\
	Sect.~\ref{sec:Approach:StudySelection:TheGathering}       & Decided data set (final result), statistics of the selection, further complementing report data \\
	\noalign{\smallskip}\hline\noalign{\smallskip}
\end{tabular}
\end{table*}

\section{Example Studies and Lessons Learned}
\label{sec:ExamplesAndLessons}
The guideline presented in this article emerges from various conducted systematic reviews and mapping studies. In this section, we provide an overview of the previously contributed studies and discuss how we applied the discussed practices and procedures so far. Table~\ref{tab:ReferenceStudies} provides an overview of the referred studies and relates the studies to the respective methods and techniques.

\begin{table}[hp]
\caption{Overview of the different studies utilizing the presented practices.}
\label{tab:ReferenceStudies}
\begin{tabular}{@{}l@{\quad}p{0.42\textwidth}@{\quad}| c | cccc | ccccc | ccc @{}}
    \hline\noalign{\smallskip}
    	\textbf{Ref.} & \textbf{Title} & \rotatebox{90}{\textbf{Type (r/m)}}  
									& \rotatebox{90}{\textbf{Preliminary Study}} 
									& \rotatebox{90}{\textbf{Trail-and-Error Search}} 
									& \rotatebox{90}{\textbf{Snowballing}} 
									& \rotatebox{90}{\textbf{Search String (1/n)}} 
									& \rotatebox{90}{\textbf{Majority Voting}} 
									& \rotatebox{90}{\textbf{Relative Rating (s/f)}} 
									& \rotatebox{90}{\textbf{Workshops}} 
									& \rotatebox{90}{\textbf{Inter-rater Agreement (s/f)}} 
									& \rotatebox{90}{\textbf{Multiple Researcher Teams}} 
									& \rotatebox{90}{\textbf{Word Clouds}} 
									& \rotatebox{90}{\textbf{Social Network Analysis}} 
									& \rotatebox{90}{\textbf{Rigor-Relevance Model \cite{Ivarsson:2011:MER:1969653.1969705}}} \\
    	\noalign{\smallskip}\hline\hline\noalign{\smallskip}
	\cite{Kuhrmann:2014vn} & A Mapping Study on the Feasibility of Method Engineering & m 
		& \cmark 
		&
		& \cmark
		& \cmark$^{(n)}$
		& \cmark$^{(3)}$
		& 
		& \cmark 
		& 
		&  
		& \cmark 
		& \cmark	
		& \\
	\cite{MOWD14} & Where Do We Stand in Requirements Engineering Improvement Today? First Results from a Mapping Study & m 
		&  
		&
		& \cmark
		& \cmark$^{(n)}$
		& \cmark$^{(3)}$
		& \cmark$^{(f)}$
		&  \cmark
		&  
		&  
		&  
		& 
		&	\\
	\cite{Kalus-G.:2013fk} & Criteria for Software Process Tailoring: A Systematic Review & r 
		&  
		&
		& \cmark
		& \cmark$^{(n)}$
		& \cmark$^{(3)}$
		& 
		&  
		&  
		&  
		&  
		&
		& 	\\
	\cite{kms2013} & Systematic Software Process Development: Where Do We Stand Today? & r 
		&  
		&
		& \cmark
		& \cmark$^{(1)}$
		& \cmark$^{(3)}$
		& 
		&  
		&  
		&  
		&  
		&
		& 	\\
	\cite{Kuhrmann:2015ix} & Software Process Improvement: Where Is the Evidence? & m 
		&  
		&
		& \cmark
		& \cmark$^{(n)}$
		& \cmark$^{(2)}$
		& \cmark$^{(s)}$
		& \cmark
		& \cmark$^{(s)}$ 
		&  
		& \cmark 
		&
		& 	\\
	\cite{Kuhrmann:2016gf} & Software process improvement: A systematic mapping study on the state of the art \emptystarmark & m 
		& \cmark 
		&
		& 
		& \cmark$^{(n)}$
		& \cmark$^{(2)}$
		& 
		& \cmark
		& \cmark$^{(s)}$ 
		&  
		&  
		&
		& 	\\
	\cite{Kuhrmann:2016ul} & How does software process improvement address global software engineering? & m/r 
		& \cmark 
		&
		& 
		& \starmark
		& \cmark$^{(2)}$
		& 
		& \cmark
		&  
		&  
		& \cmark 
		& 
		& \cmark	\\
	\cite{profes2016-SPI} & On the Role of Software Quality Management in Software Process Improvement & m/r 
		& \cmark 
		&
		& 
		& \starmark
		& \cmark$^{(2)}$
		& 
		& \cmark
		&  
		&  
		& 
		& 
		& \cmark	\\
	\cite{kmg2013a} & Towards Artifact Models as Process Interfaces in Distributed Software Projects & m/r 
		&  
		&
		& 
		& \cmark$^{(n)}$
		& \cmark$^{(2)}$
		& 
		& \cmark
		&  
		&  
		&  
		& 
		&	\\
	\cite{Theocharis:2015aa} & Is Water-Scrum-Fall Reality? On the Use of Agile and Traditional Development Practices & r 
		&  
		& \cmark
		& \cmark
		& \cmark$^{(1)}$
		& \cmark$^{(2)}$
		& 
		& \cmark
		&  
		& \cmark$^{(2)}$ 
		&  
		& 
		&	\\
	\cite{Ingibergsson:2015aa} & On the Use of Safety Certification Practices in Autonomous Field Robot Software Development: A Systematic Mapping Study & m 
		&  
		& \cmark
		& \cmark
		& \cmark$^{(n)}$
		& \cmark$^{(3)}$
		& 
		&  
		&  
		&  
		& \cmark 
		& 
		&	\\
	\cite{Racheva:2009aa} & Value Creation by Agile Projects: Methodology or Mystery? & m 
		&  
		&
		& \cmark
		& \cmark$^{(n)}$
		& \cmark$^{(3)}$
		& 
		&  
		&  
		&  
		&  
		& 
		&	\\
	\cite{5314232} & A Systematic Mapping Study on Empirical Evaluation of Software Requirements Specifications Techniques & m 
		&  
		&
		& \cmark
		& \cmark$^{(n)}$
		& \cmark$^{(4)}$
		& 
		&  
		&  
		&  
		&  
		& 
		&	\\
	\cite{Inayat2015915} & A Systematic Literature Review on Agile Requirements Engineering Practices and Challenges & r 
		&  
		&
		& \cmark
		& \cmark$^{(n)}$
		& \cmark$^{(3)}$
		& 
		&  
		&  
		&  
		&  
		& 
		&	\\
\noalign{\smallskip}\hline\noalign{\smallskip}
\end{tabular}
	Search String (1/n): The study uses 1 large or \emph{n} smaller search strings\\
	Relative Rating (s/f): Relative rating of the \underline{f}ull result set or on \underline{s}amples thereof\\
	\cmark$^{(*)}$: * number of search strings, or number of reviewers involved\\
	\emptystarmark: Study update for \cite{Kuhrmann:2015ix}; 
	\starmark: Detailed study using the dataset from \cite{Kuhrmann:2016gf}
\end{table}

\subsection{Selected Examples and Lessons Learned}
\label{sec:LL:LessonsLearned}
Over the last years of working on literature studies, we collected a number of lessons learned, which we briefly summarize below. Furthermore, in order to illustrate the lessons learned with examples, in this section, we relate the lessons learned to the studies from Table~\ref{tab:ReferenceStudies} and provide some examples. Moreover, the practices listed in Table~\ref{tab:ReferenceStudies}, in general, can be considered self-contained building blocks, i.e., they can be combined in different ways. However, in our experience, some combinations of practices showed especially beneficial. Those are presented in Appendix~\ref{sec:app:ProcedureTemplates} as a blueprint. 
We also have to note that there might exist dependencies and/or constraints providing arguments in favor or against applying certain practices in respect of a particular context (see also \cite{Zhang:2011:IRS:1968229.1968314}). For example, if a preliminary study was already conducted to find the study's scope and a set of reference publications, the ``Trail-and-Error'' search approach will not add to the study. Another example is the combination of selection strategies, i.e., the combination of majority votes, relative ratings, and voting/rating workshops. Here, setting up workshops (``expensive'' due to required human resources) should be preferably scheduled for the late selection phases when the amount of publications to be decided was reduced to a manageable number (see Sect.~\ref{sec:Approach:StudySelection:VotingProcedure}).
The rest of this section is organized according to the stages of this guideline (cf.\ Figure~\ref{fig:01:ApproachOverview}).

\subsubsection{Basic Planning}
\label{sec:LL:BasicPlanning}
Regarding the general planning activities associated with a literature study, we consider the following lessons learned the most important.

\paragraph{Make a Cunning Plan that Cannot Fail}
Given the effort, duration, and the involvement of various researchers, a literature study should be built upon a concrete plan of which the research protocol \cite{Kitchenham:2015rt} is key. We experienced that involving all researchers at the beginning is crucial to establish a shared understanding of:
\begin{itemize}
	\item The basic terms, concepts, and their synergies, in the field of interest, and
	\item The way the classification criteria should be interpreted and applied.
\end{itemize}
If classifying the relevance or other concepts based on a pre-defined scheme, those concepts need to be clarified at the beginning.

\paragraph{Watch out! The Technical Infrastructure Matters} 
One of the most important administrative tasks is to define the technical infrastructure to be used for the study. The two most important aspects are:
\begin{itemize}
	\item Use a version control system (VCS).
	\item Don't mix up Microsoft Excel and OpenOffice.org/LibreOffice.
\end{itemize}
The VCS is crucial to create baselines of the study, e.g., raw data or tentative result sets. Furthermore, a VCS allows for distributed collaborative and concurrent work, and it ensures that results are not accidentally overwritten. The second aspect is caused by practical experience: In several studies, some researchers just took the pre-configured Microsoft Excel file (see Appendix~\ref{sec:app:DataStructures}) and worked on it with OpenOffice.org/LibreOffice, so that many scripts and auto-formatting configurations did not further work, or that other researchers could simply not open it anymore with the respectively other tool (e.g., as happened in \cite{Kuhrmann:2015ix}). Fixing those situations is time-intensive and avoidable.

\subsubsection{Search Strings and Search Engines}
\label{sec:LL:SearchStringsAndEngines}
Regarding the construction of proper search strings, we consider the following lessons learned the most important ones.

\paragraph{One Search String or Multiple Ones?} Applying the introduced search strategies may result in more than one search string, which then can be customized for the different search engines. A practical problem remains: the length and complexity of the search strings, and the ability or limitations of literature databases to process search queries of and above a certain complexity (as observed when trying to replicate \cite{Schramm:2014aa}). That is, the major question is which alternative is better: One integrated long search string or multiple shorter ones, as exemplarily shown in Table~\ref{tab:SearchQueries}.
\begin{table*}[!ht]
\renewcommand{\arraystretch}{1.3}
\footnotesize

\centering
\caption{Exemplary search strings for an automated database search (excerpt from \cite{Kuhrmann:2015ix}).}
\label{tab:SearchQueries}
\label{tab:FilterQueries}
\begin{tabular}{p{0.01\textwidth}p{0.58\textwidth}p{0.3\textwidth}}
	\hline\noalign{\smallskip}
	& \textbf{Search string} & \textbf{Addresses\ldots} \\
	\noalign{\smallskip}\hline\hline\noalign{\smallskip}
	$\text{S}_{1}$ & (life-cycle \textbf{or} lifecycle \textbf{or} life cycle) \textbf{and} (management \textbf{or} administration \textbf{or} development \textbf{or} description \textbf{or} authoring \textbf{or} deployment) & process management: general life cycle \\
	& \ldots & \ldots \\
	$\text{S}_{8}$ & (feasibility \textbf{or} experience) \textbf{and} (study \textbf{or} report) & reported knowledge and empirical research \\
	\hline
	\noalign{\smallskip}\hline\noalign{\smallskip}
\end{tabular}
\end{table*}

An integrated search string has the advantage of (relative) high precision. Furthermore, it allows for capturing the entire domain in only one query. However, many literature databases, such as IEEE Xplore, have some limitations regarding length and complexity. Furthermore, the syntax of the search queries differs from database to database, thus, requiring database-specific instances of the query anyway. 
In contrast, multiple shorter search queries bypass database limitations by providing simpler structures (also recommended by \cite{Kitchenham:2015rt}) and, furthermore, those strings are easier to adapt to specific database requirements. On the other hand, in order to ensure search precision, multiple search strings require more effort in their design. For instance, to get a maximum of publications, multiple search strings require some overlap to avoid ``losses at the borders''. This, however, may cause some overhead in the result set and multiply occurring publications that have to be identified and removed later on. Furthermore, due to the simpler structure, such search strings are prone to attract unwanted publications \cite{Zhang:2011:IRS:1968229.1968314} thus requiring extra context selectors and filter constructs \cite{Kuhrmann:2015ix}.

\paragraph{Don't trust Old Result Sets}
When it comes to updating or replicating a literature study, one problem is the literature database as such. For example, in a student study activity, we aimed at replicating and updating a previously conducted SLR \cite{Schramm:2014aa} of which we had the full research protocol available. The replication package also included text files containing the database-specific search strings. In an initial test run, we encountered the following to happen: IEEE Xplore rejected the search query stating it was too complex having more than 50 terms. Transferring the (general master) search string to Scopus (to test if it will trigger any papers at all) and configuring the search properly (limiting the venues and publishers etc.), we found 215 instead of 125 papers matching the search criteria. So far, we couldn't sufficiently elaborate what happened exactly, but argue this being one of the effects coming along with continuously updating indexes (see also Brereton et al.\ \cite{Brereton:2007:LAS:1225950.1226109}, who mention indexing of current digital libraries inappropriate). In short, over the time, search queries age and literature databases evolve. There is no guarantee that a result set obtained at one point in time will be re-constructible some time later. 
There is no mitigation strategy for this problem, except to increase the transparency of the data collection by reporting a timestamp for the searches to support the reproducibility and thereby the validity.
Therefore, search queries as well as raw result sets (Sect.~\ref{sec:Approach:DataCollectionCleaning}) should be stored---at least to reproduce the findings from the raw data.

\subsubsection{Data Collection and Cleaning}
\label{sec:LL:DataCollectionAndCleaning}
Regarding the data collection and cleaning, we consider the following lessons learned the most important ones.

\paragraph{Find the Right Scope}
In some studies, we saw an explicit and intentional limitation of the search; for instance, instead of searching a whole library, authors of a study limited themselves to particular conferences or journals \cite{Schramm:2014aa}. Such an approach promises the advantage of having a more focused result set by avoiding overhead \cite{Kitchenham:2015rt}. However, this may come possibly at the price of  information loss, because many relevant publications might not be found. Such procedure is of course possible, but not recommended; yet, if conducted that way, it should be explicitly mentioned in the threats to validity to increase the transparency and reproducibility. Finally, if the ultimate goal is a systematic mapping study, however, this approach cannot be applied, as the limitation of the search scope hampers the overall result set quality and also the quality and reliability of the conclusions.

\paragraph{What Publication Type to Include?}
Besides the used search engines, researchers need to clarify what types of publications can/cannot be included into the result set. We consider, for example, including textbooks and edited chapters as a viable option in case the study is about the analysis of definitions, e.g., to understand the meaning of a particular concept as used by authors in a field. The choice of certain books can and should be justified based on their popularity in a community; for example by including well-established textbooks as used for teaching, or books that have a high number of citations in empirical papers in the area. Master theses in turn should be avoided given their missing peer-review process. Involving Ph.D.\ theses, however, depends on various contextual characteristics; for instance, whether they passed a peer-review process or whether they are cumulative ones (which might, of course, lead to duplicates in the result set given that the content is previously published material, see also Sect.~\ref{sec:Approach:DataCollectionCleaning:Collection}).

\paragraph{How Valid is the Paper Selection Process?} 
In the previous sections, we described different voting procedures that can be applied. With every voting procedure comes different ways of increasing the validity of the methods applied and the results obtained. The least common denominator of all procedures, however, is the inter-rater agreement \cite{Kitchenham:2015rt}. We postulate the use of inter-rater agreements especially if used in a multi-staged voting procedure as they serve as a constructive quality assurance measure between the stages; for example, to clarify misconceptions, misinterpretations of research questions, misinterpretations of classification schemes, and different understanding of the relevance of publications. Besides the value of inter-rater agreements for constructive quality assurance, it also increases the transparency to the reader and, therefore, the conclusion validity. However, such an agreement makes only sense if the voting procedure is not conducted iteratively over incomplete result sets whereby it is impossible to use the agreement as a means to improve the classification between stages (if not used in a training/test phase). Hence, there is a trade-off regarding the purpose and the effort of using the inter-rater agreement, which needs to be clarified in advance. 

\paragraph{How Much is Enough?}
As a matter of fact, there is no meaningful metric that could be used to indicate whether the result set is sufficiently large or not, let alone because the size of a dataset provides no indicator to the quality of its content \cite{Wohlin20132594,Zhang:2011:IRS:1968229.1968314,Badampudi:2015:EUS:2745802.2745818}. For example, in \cite{Kalus-G.:2013fk} and \cite{Ingibergsson:2015aa}, we performed the data search, but then capped the result sets to include only the first 50 hits per query result. Is this enough? What is the risk of loosing relevant papers? As there is no common ground, such a decision must be taken on a per-study basis. Yet, it needs to be ensured that the result set of papers obtained is of high quality, i.e., representative for the field of investigation and the research questions formulated. This means to ensure an accurate result set and a detailed and validated review protocol including a search string potentially adapted to the particularities of the search engines, and detailed inclusion and exclusion criteria.

\subsubsection{Preparing the Handover}
\label{sec:LL:PreparingTheHandover}
Although a study selection might be completed, more activities can be carried out before entering the in-depth analysis. The final dataset provides already data that can be used early in the overall literature study process to help researches finding appropriate points to start with the analysis. From our so far conducted studies, we consider the following lessons learned helpful.

\paragraph{Exporting Keyword Lists, Abstracts, and Word Clouds}
From the result set, keyword lists and abstracts can be easily harvested and prepared to support the beginning of the analysis. We can create, for example, word clouds from these lists to get a quick visual inspection where a striking keyword could indicate to a set of publications to start with.
However, what seems easy to generate and use can eventually turn out to be difficult or even misleading: several tools for word cloud generation have limitations regarding the amount of text they can process. A solution is to perform a keyword coding, which serves three purposes (as used in \cite{Kuhrmann:2015ix}): first, the list of keywords is shortened; second, the used terminology is harmonized (e.g., ``small-to-medium-sized companies'' and ``small and medium enterprises'' are coded to ``SME''); third, the keyword coding can be considered a first step towards full coding, which is normally performed in the context of a mapping study to work out the classification schemas. If a keyword (and/or abstract) coding would be performed, the outcomes of the activities would comprise the respective keyword lists, abstract lists, the mapping files containing the codes and all synonyms, and optionally generated visuals.

\paragraph{Utilizing Social Network Analysis as a Means for Pre-Selection}
A social network is a graph that provides an overview of subjects and their relationships (see for instance \cite{Hanneman2005,Scott2010,Wasserman1994}). Right in the early stages, even before the actual study begins, a social network graph can be generated from the result set. Such a graph can serve multiple purposes. For instance, a social network graph highlights cooperation cliques, i.e., authors that collaborate and contribute a considerable share of the result set, thus, forming the ``community leaders''. When it eventually comes to begin with the result set analysis, researchers can face the problem to find a proper starting point. Potentially identified clusters can provide some guidance through the result set. Another option is to look for domain-shaping key contributions, which are potentially highlighted by a citation network\footnote{This approach needs to be considered with care, as for instance newer publications may have a high-quality contribution, but don't have a high citation count (e.g., compared to a 10-year old publication). Therefore, citation networks only deliver initial indication and trends shouldn't be taken for granted.}.
Beyond the analysis preparation, a social network is also a supportive means within a study. For example, in \cite{Kuhrmann:2014vn}, we used a collaboration network to study if a found trend in the publication space is just because of the result set's background noise. Therefore, we generated the social network to identify the key contributors and created a sub-map, which was based on the respective publications only, and compared whether the general trends differed.

\section{Related Work and Discussion}
\label{sec:RelatedWork}
This article complements a number of existing guidelines and initiatives for conducting literature studies. In this section, we  provide an overview of related work including approaches, methods, experiences, and tools to support literature studies and position our contribution in context of the current publication landscape. Table~\ref{tab:ApproachComparison} summarizes the body of knowledge in existing guidelines we found particularly relevant and adds how our contribution at hand deviates (i.e., adapts/extends) from existing contributions.
\begin{table*}[!ht]
\renewcommand{\arraystretch}{1.3}
\footnotesize

\centering
\caption{Relation of the present guideline with further established guidelines.}
\label{tab:ApproachComparison}
\begin{tabular}{lp{0.5\textwidth}p{0.6\textwidth}}
	\hline\noalign{\smallskip}
	\textbf{Ref.} & \textbf{Key Contributions} & \textbf{Adaptation/Extension} \\
	\noalign{\smallskip}\hline\hline\noalign{\smallskip}
	 \cite{Kitchenham:2015rt} & Kitchenham et al.\  provide a well elaborated overview of the systematic literature study processes. To this end, they introduce a conceptual description of what to do in a systematic review or in a systematic mapping study, and an explanation of why these steps should be carried out. The aim is to provide a generalized view on \emph{what to do} while concrete advice of \emph{how to operationalize} the respective steps in a specific context is out of scope. & Our guideline emphasizes the operationalization of the particular steps in the data collection and study selection phase, and the guide provides examples and critical discussion of lessons we learned. Furthermore, our guideline describes activities as building blocks and offers exemplary workflow templates for literature studies of different complexity and size. \\
	\cite{PVK15} & 	 Peterson et al. propose a guideline, which extends their original one \cite{PFMM08} grounded in evidence obtained from analyzing 52 mapping studies and comparing the guidelines used therein. The guideline provides a checklist of activities and refers to articles that used those to select data for the study. It further proposes a more detailed classification schema (compared to \cite{PFMM08,Wieringa:2005:REP:1107677.1107683}) and comprises small examples for illustration. & Our guideline has a different scope compared to \cite{PVK15} as we focus on the relatively unexplored early stages only. That is, our guideline focuses on the data collection and study selection process, whereas we pay little attention to the data extraction and analysis which we believe to be already well elaborated. Yet, our guideline provides a more detailed perspective, e.g., on the different practices and how to combine them, how to utilize techniques such as word clouds or social networks to aid the selection process (both not mentioned in \cite{PVK15}). Therefore, our guideline is a pragmatic complementation of the \emph{study identification} phase from \cite{PVK15}. \\
	 \cite{Zhang:2011:IRS:1968229.1968314} & Zhang et al.\ describe a ``quasi-gold standard'' to find an effective study selection strategy. Among other things, Zhang et al.\ define a search process to achieve high sensitivity and precision of the searches. &  Similar to Zhang et al., our guideline recommends utilizing different search engines. Yet, our guideline provides more details regarding actual practices to analyze and clean the result (sub-)sets obtained from different search runs, and we also provide recommendations to develop an integrated result set to be evaluated in the actual study selection process. Therefore, our guideline complements \cite{Zhang:2011:IRS:1968229.1968314} and provides recommendations to fill gaps, such as missing information concerning the steps required to get from step 4 (conduct automated search) to step 5 (evaluate search performance). \\
	 \cite{Wieringa:2005:REP:1107677.1107683} & The work by Wieringa et al.\  has become representative for developing classification schemas based on a well elaborated reference (see also \cite{PFMM08,Paternoster20141200} or \cite{PVK15}). & In the present guideline, we explicitly do not aim to support schema development. However, when providing a data structure template, we leave room for classification schemas. Furthermore, grounded in our experience, we also propose considering free metadata to be collected, since we found strict classification schemas not well-applicable in all setups.  \\
	 \cite{Ivarsson:2011:MER:1969653.1969705} & The rigor-relevance model by Ivarsson et al. provides a scale-based approach to determine the relevance to industry and the rigorousness of the research conducted. Hence, this model can support the paper selection process.  & In our guideline, we utilize the rigor-relevance model exactly as proposed as an explicit extra dimension to support the classification, because we experienced it to be of particularly high value. We therefore recommend to use a combination of ``standard schemas'' (e.g., \cite{Wieringa:2005:REP:1107677.1107683,Ivarsson:2011:MER:1969653.1969705,PFMM08,Paternoster20141200}) complemented with study-specific schemas, e.g., those developed from free metadata.  \\
	\noalign{\smallskip}\hline\noalign{\smallskip}
\end{tabular}
\end{table*}

\paragraph{Approaches}
We deliberately use the term ``approaches'' to subsume all the different processes and methods utilized in literature studies. One prominent approach in context of literature studies is the \emph{systematic review} process as initiated for software engineering by Kitchenham \cite{Kitchenham:2004fk} and continuously improved, e.g., Kitchenham and Charters \cite{Kitchenham:2007fk}, eventually leading to a consolidated guideline \cite{Kitchenham:2015rt}, as well as the \emph{systematic mapping study} made popular for software engineering by Petersen et al.\ \cite{PFMM08} (updated in \cite{PVK15}). 

These general guidelines, which---despite of their value to provide a common structure and consistent terminology---have been experienced as too generic for direct practical application~\cite{PVK15,Staples:2007:EUS:1282869.1282969}. They still serve as an umbrella and a multitude on fine-grained methods and models, and advice and best practices can be embodied by the guidelines. For example, a challenge in literature studies is the development of proper classification schemas. In literature, we find, for instance, the \emph{research type facet} classification schema developed by Wieringa et al.\ \cite{Wieringa:2005:REP:1107677.1107683} and the \emph{contribution type facet} schema as illustrated by Petersen et al.\ \cite{PFMM08} (adopted from Shaw's work \cite{Shaw:2003:WGS:776816.776925}) serving as generic classification patterns for studies \cite{PVK15}. Another perspective is provided by Paternoster et al.\ \cite{Paternoster20141200}, who utilize a \emph{focus type facet} and a \emph{pertinence facet}. Furthermore, Paternoster et al.\ \cite{Paternoster20141200} include a model for determining \emph{rigor and relevance} of the involved studies (based on a model proposed by Ivarsson and Gorschek \cite{Ivarsson:2011:MER:1969653.1969705}) to support the determination of the result set's reliability. However, Petersen et al.\ \cite{PVK15} mention those classification schemas critical. The reason is that such schemas, as the one by Wieringa et al.\ \cite{Wieringa:2005:REP:1107677.1107683}, leave room for interpretation. As a matter of fact, we can find ``tailored'' variants of this schema in a number of studies (see also Wohlin et at.\ \cite{Wohlin20132594}). It also remains a challenge to construct a schema in a proper and efficient manner, and a number of strategies are available for this purpose \cite{6092586}. For instance, in our study \cite{Kuhrmann:2015ix}, we used the \emph{focus type facet} concept finding the described construction procedure from \cite{Paternoster20141200} inappropriate for the following reasons: if one has to deal with a very large number of papers, a manual coding-based schema construction is too costly. Moreover, it is challenging to clearly define the elements of such a schema, as indicated by Portillo-Rodr\'{i}guez et al.\ \cite{PortilloRodriguez2012663}. This is because not all papers have sufficient information in title, keywords, and abstracts to conduct a proper and fine-grained classification \cite{Brereton:2007:LAS:1225950.1226109}, and if the purpose of the study is to capture an entire domain, developing a precise classification is close to impossible, as many publications address multiple topics, which makes a unique classification hard or even impossible. Therefore, in previous work \cite{Kuhrmann:2015ix}, we started collecting ``free'' metadata instead of providing a big picture of the domain, but leaving the full classification to the fine-grained analyses of selected topics. As outcome, in \cite{Kuhrmann:2016gf,Kuhrmann:2016ul} we used the metadata to generate \emph{heat maps} (as also done in \cite{Penzenstadler:2014:SMS:2601248.2601256}) to work out trends worth further investigation. 

Constructing a classification schema requires data to apply the schema. In this respect, Petersen et al.\ \cite{PVK15} found 15 ways to collect and identify relevant studies. Data search is mainly done using manual and database searches, and snowballing. Yet, it is currently subject to discussion which of the practices (or combinations thereof) result in datasets of sufficient quality \emph{and} what is considered a sufficient dataset after all \cite{Wohlin20132594}. Ali and Petersen \cite{Ali:2014:ESS:2652524.2652557} review strategies to select studies in systematic reviews and formulate a selection process. They conclude that a good-enough sample could be obtained by following a less inclusive but more efficient strategy. Zhang et al.\ \cite{Zhang:2011:IRS:1968229.1968314} present a ``quasi-gold standard'' to identify relevant studies and Badampudi et al.\ \cite{Badampudi:2015:EUS:2745802.2745818} show that snowballing also leads to an appropriate result set. That is, all the different search strategies used so far produce sufficient datasets. Up to now, however, little has been reported on the complementary use of the different search strategies, costs and benefits associated with such a combination. In the present article, similar to Dyb\aa\ et al.\ \cite{Dyba:2007:ASR:1302496.1302940}, we stress this aspect by presenting the combined use, and we also demonstrate how a search can be complemented by further techniques, such as \emph{social network analysis} \cite{Scott2010,Wasserman1994} or \emph{word clouds} \cite{Kuhrmann:2015ix,Kuhrmann:2016ul}, to support pre-selection, analysis scoping, and dataset/result visualization. 

The search and selection procedures also include the definition and use of inclusion and exclusion criteria. However, Petersen et al.\ \cite{PVK15} found only five out of 10 guidelines explicitly addressing this topic, but there was so far no attempt to craft a set of standard in-/exclusion criteria. Similarly to standard research questions, standard data collection workflows, and standard study selection procedures, we have proposed a set of standard inclusion and exclusion criteria to support a quick start of the study and to lay the foundation for the development of further study-specific criteria.

\paragraph{Experiences}
Regarding the (generic) guidelines used by empirical software engineering researchers, Petersen et al.\ \cite{PVK15} found and compared in total 10 guidelines used, whereas the (more general) ones by Kitchenham and Charters \cite{Kitchenham:2007fk} and Peterson et al.\ \cite{PFMM08} were identified as the most frequently used. Furthermore, their findings include identified gaps in the individual guidelines, such as missing practical advice on how to do self-evaluation, justification and motivation of the research question chosen regarding the demographic overview of a planned study, or missing shared practices from personal accounts of designing systematic reviews and mapping studies by following specific guidelines. Petersen et al.\ \cite{PVK15} add to a series of meta-studies that aim to monitor the guidelines' application and to collect lessons learned and best practices is a required step to consolidate experience. For instance, Kitchenham and Brereton \cite{Kitchenham:2013:SRS:2535048.2535149} analyzed 68 studies and found that the time required to conduct a systematic review and difficulties regarding quality assessment are problematic. This finding provides extra arguments for sophisticated tool support. In their study, authors also found current digital libraries not appropriate for broad literature searches. This is also supported by Brereton et al.\ \cite{Brereton:2007:LAS:1225950.1226109}, who specifically found the indexing of those digital libraries inadequate and also mention that the quality of paper abstracts is too poor, e.g., to judge upon the relevance of a paper based on its abstract only. This provides a rationale for different search and selection strategies \cite{Ali:2014:ESS:2652524.2652557,Badampudi:2015:EUS:2745802.2745818,Zhang:2011:IRS:1968229.1968314}. A more general discussion is raised by Staples and Niazi \cite{Staples:2007:EUS:1282869.1282969}, who generally recommend using guidelines, but also mention a need to optimize the process as such (e.g., narrowly defined research questions, improved selection procedures, and improved data extraction) to reduce the effort needed to conduct such a study. However, exemplary research questions to start a literature study are only provided by Petersen et al.\ \cite{PVK15} as part of the analysis of other studies, thus, being focused to the respective study subjects---the presented list of quoted research questions does not serve the generalization. Dyb\aa\ et al.\ \cite{Dyba:2007:ASR:1302496.1302940} consider ``normal'' meta-analytic approaches to be of limited use for software engineering only and, hence, report their experience from applying diverse study types in a systematic review; a mixed-method approach similar to the practices reported in the present article. Riaz et al.\ \cite{Riaz:2010:ECS:2227057.2227063} provide a different perspective in their report and mention experts and novices having a different perception of the systematic review process and its challenges. The present article also addresses this point by providing examples, reusable assets like research questions or in-/exclusion criteria, and a detailed elaboration on selected practices and a demonstration of their use. Such challenges are also addressed by Fabbri et al.\ \cite{6680572}, who provide an experienced-based guideline that comes as integrated process with the purpose of externalizing tacit knowledge about the process and its implementation. In contrast to Fabbri et al.\ \cite{6680572}, the present article is not supposed to be a self-contained comprehensive guideline covering the process of conducting a literature study as a whole. Instead, we focus on the early stages and provide a limited, but interlinked set of practices illustrated by examples and reusable building blocks, which we also compile into reference workflows to follow.

\paragraph{Tools}
The body of knowledge in software engineering is growing and, thus, literature studies are likely to grow in size and complexity as well. Tool support has therefore become crucial to collect, manage, and evaluate data. However, the question of what can be considered as proper tool support has puzzled researchers for years \cite{Hassler2016122,Tell:2016rz}. A group around Marshall conducted research on tool support for literature studies \cite{6681371,Marshall:2015:SRT:2745802.2745824,Marshall:2014:TSS:2601248.2601270,Marshall:2015:TSS:2745802.2745827}. Among others, they provided a feature analysis to define basic requirements \cite{Marshall:2014:TSS:2601248.2601270}, and in \cite{6681353}, authors found a strong need to provide support for planning and teamwork when conducting a literature study. In \cite{Marshall:2015:TSS:2745802.2745827}, the same author group concluded a recommended list of requirements, which was generated based on 13 semi-structured interviews. Yet, the requirements list only provides a high-level overview of features that opens a fairly large design space that should be carefully considered when designing tools. The challenges coming along with this large design space were explicitly addressed in \cite{Tell:2016rz} in which we, based on a shared set of requirements, independently developed two tools---both realizations with different features emphasized and implementing different work and collaboration patterns. Over the years, few tools dedicated to support researchers performing systematic reviews have been proposed; notable examples are SLuRp \cite{Bowes:2012:STH:2372233.2372243}, SESRA \cite{Molleri:2015:SWA:2745802.2745825}, and StArt \cite{hernandes2012start}.  These tools were analyzed in \cite{Marshall:2014:TSS:2601248.2601270}, yet those are not ranked with flying colors. Still, the classic spreadsheet application (quite often) in combination with so-called reference managers (e.g., EndNote, Mendeley, Papers, and Zotero) seem to build the standard tooling for literature studies.

\paragraph{Summary of Related Works}
The present article contributes to the body of knowledge by stressing the need for more concrete advice to complement the generic guidelines, and by offering an experience-based guideline especially to perform the steps in the early stages. Although for instance Petersen et al.~\cite{PVK15} provide a comprehensive selection of practices used for these stages, a streamlined approach to presenting, explaining, and linking these steps to each other is not in scope of their contribution. In a nutshell, most of the available guidelines are focused on \emph{what} a design should accomplish rather than on \emph{how} and \emph{why} a particular step should be executed in a cost-effective way. For example, we found no guideline explaining what pieces of information are worthwhile including and what justified particular configurations of descriptive data pieces to be taken care of by the researchers. Our recommended minimal data structure (Table~\ref{tab:StandardDataCollectionDataStructure}) can be directly used by researchers facing this question. Furthermore, no guideline so far discussed in detail the ways to run a voting procedure. We provided an operationalized description on how to do this in a systematic way, along with a discussion on a research team model, a scaling/vote calculation schema and a demonstration of a potential technical realization based on the suggested minimal data structure (Figure~\ref{fig:ColorCodedSpreadsheet}). Based on our reported experience, we also provide a description of the work deliverables that are produced during a literature study process and the dependencies among the deliverables (Sect.~\ref{sec:Approach:Concluding}), and we shared our lessons learned regarding the issues coming along with handling search engines, which are barely discussed in available guidelines.

\section{Conclusion}
\label{sec:Conclusion}
Systematic literature studies have become a powerful means to elaborate and structure the state of reported knowledge. Especially in the software engineering community, they have received much attention in recent years. Despite their relevance to the community and first valuable proposals of guidelines, they are difficult to conduct, require a lot of effort and depend on experiences and expertise of the researchers involved. Especially the latter decides often over the success of a study, depending on aspects such as
\begin{itemize}
	\item Appropriateness of the research questions and value to the community,
	\item Accuracy of the design, or
	\item Reproducibility of the data collection.
\end{itemize}
When conducting literature studies, there are various challenges all concerning the initial stages of the data collection rather than the particularities of the later analytical phase, and there are challenges that concern the organization of such a study. 

In this article, we reported on our experiences made in the course of various literature studies and contributed an experience-based guideline that puts strong emphasis on tackling some practical challenges. Our aim was to specifically support young scholars facing their first literature study and to provide them with a pragmatic and easy-to-enact guideline. To this end, we collected and structured our experiences, and we also shared our experiences in utilizing different tools to support the data collection, the dataset cleaning, or the study selection procedures. Furthermore, we provided some generalized blueprint-style workflows to follow in a particular study, also increase the efficiency in the way study designs are reported in papers within the space limitations given for conference submissions so that the used approaches don't have to be justified from scratch all the time drowning the presentation of the results out.

While compiling this guideline, we also realized again the need for fine-grained guidelines and, moreover, the need for a sophisticated tool support. As a matter of fact, all our studies were conducted utilizing fairly simple tools, such as spreadsheets or plain text files to feed further external tools, e.g., word cloud generators. However, having conducted the data collection, research teams have rich data available, which could be used for extensive tool-support. Yet, comprehensive tools are not yet publicly available or, if at all, in their early stages of their development as for instance \cite{Tell:2016rz}. This indicates to a strong need to (1) increasing the effort spent on developing applicable procedures and fine-grained reference workflows from the available knowledge and experience, and (2) to put effort into the development of tools to support literature studies. These tools need to support the collection of data, their storage and organization, the management of in-/exclusion criteria, support to implement workflows for paper selection and classification, which also includes the management of classification schemas, and, eventually, supporting the connection to further tools, e.g., word cloud generators, statistics software, and social network analysis tools. 

\subsection*{Acknowledgements}
We want to thank Roel Wieringa for fruitful discussions on previous versions of this article and all our students, especially those who contributed to our previously conducted literature studies over the past years. Finally, we are grateful for the constructive feedback provided by the anonymous reviewers of this article who helped improving it substantially.  

\appendix

\section{Study Workflow Templates}
\label{sec:app:ProcedureTemplates}
In this appendix, we provide selected workflow templates, which we inferred from experiences (Table~\ref{tab:ReferenceStudies}) for simple reuse in research method descriptions of scientific papers. The provided templates can be used to inspire or shorten the description of research methods, which especially in conference papers consumes much precious space. For each model in subsequent sections, we provide a brief context description, an exemplary workflow, and textual description.

\subsection{Template 1: 2 Researcher Workshop Model with Snowballing}
\label{sec:app:2ResearcherWS-SB-Model}
\begin{figure*}[!ht]
  \includegraphics[width=0.75\textwidth, right]{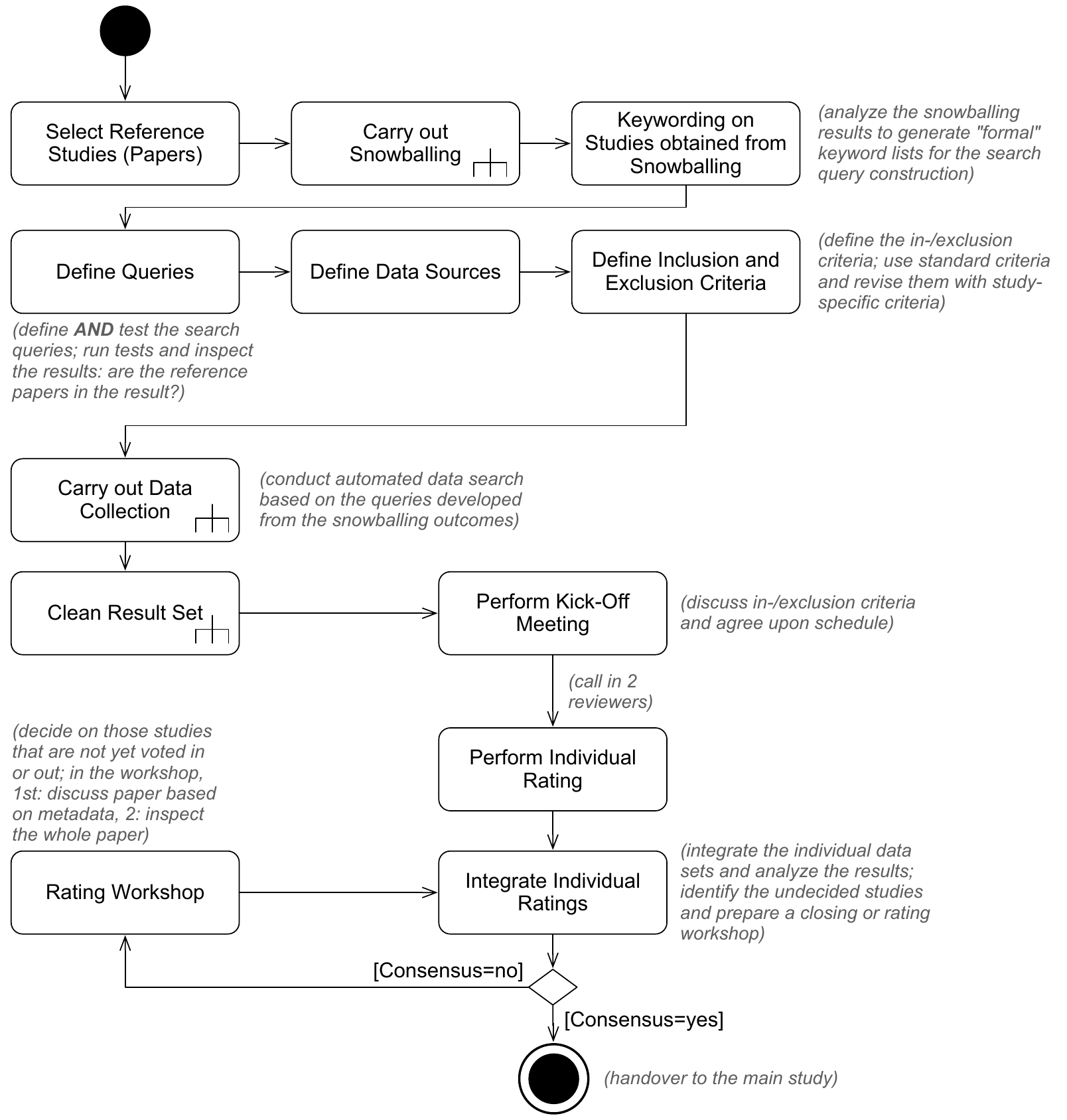}
  \caption{Exemplary workflow for the 2 researcher workshop model with a snowballing-based preliminary study.}
  \label{fig:2rws-model}
\end{figure*}

\textbf{Context:} This model addresses smaller literature studies in which just two researchers collaborate, thus, having no option to implement more comprehensive study selection procedures, such as majority votes. Our experience shows this model to be well-applicable in settings with up to approximately 50 papers, two senior or one senior and one junior researcher, and in distributed settings. Apart from an initial research objective and/or a set of research questions and a (small) set of reference publications, no extra entry conditions need to be fulfilled.
\\[0.5em]
\textbf{Workflow:}
Figure~\ref{fig:2rws-model} illustrates the basic workflow for this model including some notes emphasizing the most relevant points to be considered.
\\[0.5em]
\textbf{Workflow Description:}
The \emph{2 Researcher Workshop Model with Snowballing} is implemented as follows: Right in the beginning of the study, a snowballing-based preliminary study is conducted. For this pre-study, a set of reference papers is selected to lay the foundation for an (incremental) snowballing search. When the snowballing is done, the obtained papers are analyzed for keywords, which are used to construct the search queries for an automated database search. As the last preparation steps, the data sources of interest are selected and the inclusion and exclusion criteria are defined.

The data collection is performed (according to the search strategy, Sect.~\ref{sec:Approach:DataCollectionCleaning:Collection}). After the search, the dataset is cleaned (Sect.~\ref{sec:Approach:DataCollectionCleaning:Cleaning}), e.g., by a stepwise integration of individual datasets. The kick-off meeting is---on the one hand---closing the data collection and cleaning phase and---on the other hand---starts the study selection phase. In the kick-off meeting, both researchers reflect on all the criteria, inspect and prepare the dataset for the rating, and agree on a schedule. According to the procedure illustrated in Figure~\ref{fig:05:StandardMajorityVoting}, each researcher gets a copy of the dataset and carries out the individual rating. When the rating is done, both datasets are integrated and checked for consensus. In a rating workshop (or multiple workshops), both researchers iterate through the dataset discussing all items that are not yet decided to find an agreement. When the concluding integration is done, the study selection phase is closed and the result set is transferred to the main study (Sect.~\ref{sec:Approach:Concluding}). For handing over the result set, a copy of the fully rated result set is created for archiving, and the actual result set is reduced, i.e., those dataset items that were rated as irrelevant for the main study are removed from the dataset so that only relevant data finds its way into the analysis.

\subsection{Template 2: 3 Researcher Voting-only Model}
\label{sec:app:3ReseaercherVotingOnly}
\begin{figure*}[ht]
  \includegraphics[width=0.75\textwidth, right]{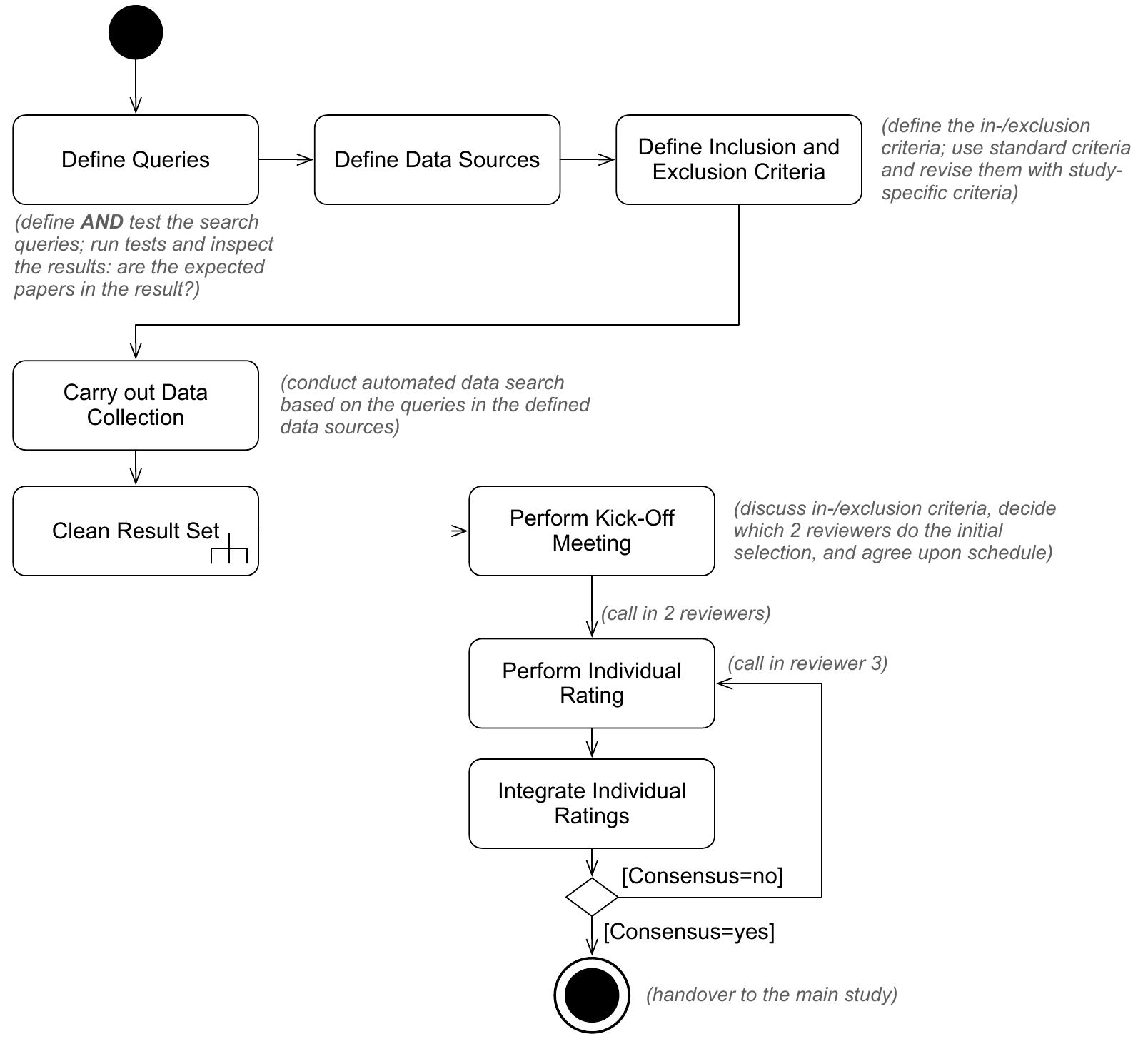}
  \caption{Exemplary workflow for a data collection and study selection approach for 3 reviewers using a voting-only approach.}
  \label{fig:3rvoting-model}
\end{figure*}

\textbf{Context:} This model addresses literature studies in which three researchers collaborate and implement a voting-based study selection procedure. Our experience shows this model to be well-applicable in the majority of all literature study settings. This model supports mixed and distributed teams, whereas at least one senior researcher has to be involved to guide the study project. Our standard implementation of the \emph{3 Researcher Voting-only Model} follows the 2+1 approach (Figure~\ref{fig:05:StandardMajorityVoting}, p.~\pageref{fig:05:StandardMajorityVoting}), i.e., the voting procedure to select relevant papers is organized by two researchers carrying out the full voting independently and calling in a third researcher to make the final decisions. In order to set up a study following this model, research objectives and questions, keyword lists and accordingly derived search queries have to be in place; optionally, a (small) set of reference publications is available.
\\[0.5em]
\textbf{Workflow:}
Figure~\ref{fig:3rvoting-model} illustrates the basic workflow for this model including some notes emphasizing the most relevant points to be considered.
\\[0.5em]
\textbf{Workflow Description:}
The \emph{3 Researcher Voting-only Model} is implemented as follows: After defining the search queries, data sources of interest, and the required inclusion and exclusion criteria, actual data collection is performed (Sect.~\ref{sec:Approach:DataCollectionCleaning:Collection}). After the data collection, the data sets are cleaned (Sect.~\ref{sec:Approach:DataCollectionCleaning:Cleaning}), e.g., via a stepwise integration of individual datasets.

In the kick-off meeting, the team of researchers nominates two researchers who will conduct the initial rating. According to the procedure illustrated in Figure~\ref{fig:05:StandardMajorityVoting}, each of the two selected researchers gets a copy of the integrated dataset for carrying out the individual rating. When both researchers have rated the dataset, one of them integrates both and analyzes the integrated result set for the agreement. Those dataset items that are not yet decided are selected and exported in a reduced dataset, which is given to the third reviewer. The third reviewer then performs a rating on the reduced dataset and, eventually, integrates the outcome with the full dataset. After performing this third rating, the dataset is now fully decided and can be prepared to be transferred to the main analysis (Sect.~\ref{sec:Approach:Concluding}). If using a tool-supported approach as, for instance, shown in Figure~\ref{fig:ColorCodedSpreadsheet}, the different stages can be supported by simple calculation, scripts, and conditional formatting (color coding).

\section{Recommended Data Structure}
\label{sec:app:DataStructures}
\begin{figure*}[!ht]
  \includegraphics[width=\textwidth, right]{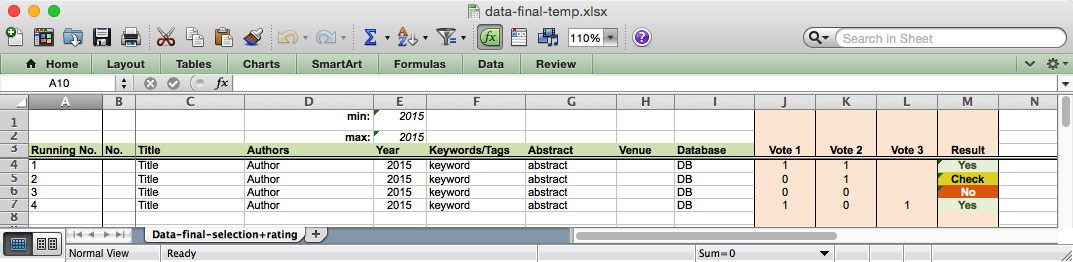}
  \caption{Example of a color-coded voting spreadsheet. The sheet shows different combinations of a 3-person majority vote (2 reviewers + 1 extra reviewer for final decisions).}
  \label{fig:ColorCodedSpreadsheet}
\end{figure*}

In this section, we present a recommendation of a data structure to store data obtained by a manual/automatic literature search. Table~\ref{tab:StandardDataCollectionDataStructure} presents this recommended data structure, which emerges from several literature studies (Table~\ref{tab:ReferenceStudies}), and the table explains the meaning of the different fields. Note: We consider the presented data structure to be \emph{minimal}, i.e., specific studies will require further data fields. However, due to the absence of comprehensive and mature tools to support mapping studies, the normal would be to set up a simple spreadsheet. Examples of such spreadsheets (Figure~\ref{fig:ColorCodedSpreadsheet}) can be obtained from \url{http://goo.gl/PBylsn}.
\begin{table*}[!htb]
\centering
\caption{Recommended minimal data structure.}
\label{tab:StandardDataCollectionDataStructure}
\begin{tabular}{@{}p{0.12\textwidth}l@{\quad}p{0.675\textwidth}@{}}
    \hline\noalign{\smallskip}
    	\textbf{Field} & \textbf{Cardinality} & \textbf{Description}  \\
    	\noalign{\smallskip}\hline\hline\noalign{\smallskip}
	No. 						& 1 		& The overall publication number in the integrated dataset. \\
	DB-No. 					& 1 		& The database-specific number if a paper from the individual literature database to allow for linking an entry to the originating dataset. \\
	Title 						& 1 		& Title of the publication. \\
	Authors 					& 1, 1..n 	& Authors of the publication; either integrated in one cell and separated by special characters (e.g., ``;''), or converted into a one-author-per-cell pattern, i.e. there are n columns to represent the author list. \\
	Keywords 					& 1 		& List of keywords separated by special characters (e.g., ``,'' or ``;''). \\
	Abstract 					& 1 		& Abstract of the paper. \\
	Year 						& 1 		& Year of publication (note: e.g., for journals, there might be multiple dates, such as accepted, online available, preprint, published, etc.---it is required to define which of these is the one that makes it into the dataset). \\
	Publisher/ Database 				& 1		& Which database created this item? In case of cross-indexing, publisher and originating database can differ, e.g., IEEE Xplore also lists IET papers. \\
	Source/ Venue 				& 1 		& Which source or venue published this paper? In case of a conference, this field should contain the conference name and/acronym, in case of a journal, the name/acronym of the journal should be contained, and so forth. \\
	Publication Vehicle 			& 1..n 	& For every publication vehicle, an individual column should be present, e.g., journal, magazine, conference, workshop, book, chapter, misc, and so forth. Experience shows individual columns beneficial for later analyses. \\
	General Comments 			& 1 		& Provide some space for general comments. \\
	Metadata Classes (optional)	& 0..n	& It was shown beneficial to provide some space for metadata, for example, this is a survey, a literature review, this deals with Agile, and so forth. The number of metadata is not limited and can be extended during analysis. Furthermore, metadata should allow for categorization, that is, one column per metadata class should be provided. \\
\noalign{\smallskip}\hline\noalign{\smallskip}
\end{tabular}
\end{table*}

The data structure as presented in Table~\ref{tab:StandardDataCollectionDataStructure} only contains a minimal set of data, which needs to be extended according to the study's scope. For systematic mapping studies, the following extra data should be contained:
\begin{itemize}
	\item Generic/reused classification schemas, such as research/contribution type facet (Wieringa et al.\ \cite{Wieringa:2005:REP:1107677.1107683}, Petersen et al.\ \cite{PFMM08})
	\item Study-specific classification schemas, such as focus type facets (Paternoster et al.\ \cite{Paternoster20141200}) or rigor/relevance models (Ivarsson and Gorschek \cite{Ivarsson:2011:MER:1969653.1969705})
	\item In-/exclusion criteria to document, why a paper was in-/excluded (cf.\ Table~\ref{tab:StandardInExclusionCriteria})
\end{itemize}
Furthermore, grounded in our experience from \cite{Kuhrmann:2015ix}, we also recommend adding ``dynamic metadata'' to the data structure (as already mentioned in Table~\ref{tab:StandardDataCollectionDataStructure}). Such metadata can be added on-the-fly and can support the enhancement of the dataset. From our experience \cite{Kuhrmann:2016gf}, we recommend to collect metadata at least from the dimensions \emph{Study} and \emph{Context}.

The dimension \emph{Study} covers the overall research approach followed in a particular paper, e.g., is a particular paper a primary study, a replication, or even a secondary study, and it can even contain the research methods used, such as interview research or grounded theory analyses. Metadata from this category supports a more detailed classification and analysis of papers regarding the research and contribution type facets. The dimension \emph{Context} aims at collecting as much context information from the selected papers as possible, such as the software engineering lifecycle phase addressed by a paper (e.g., design, coding, test), the organizational context in which the research was conducted (e.g., SMEs, global players etc.), and the application domain of a paper (e.g., automotive software or software for the healthcare domain).

\bibliographystyle{spmpsci}      
\bibliography{sms-method}
\end{document}